\documentstyle[times,graphics,epsfig,amsmath,amssymb,natbib,psfrag,color]{mn2e}

\newcommand{\be}{\begin{equation}}
\newcommand{\e}{\end{equation}}
\newcommand{\bear}{\begin{eqnarray}}
\newcommand{\ear}{\end{eqnarray}}

\def\xh1{x_{{\rm H~{\sc i}\,}}}
\def\aap{AAP}
\def\apj{ApJ}
\def\aj{AJ}
\def\apjs{ApJS}
\def\apl{ApJL}
\def\mnras{MNRAS}

\def\nat{Nature}
\def\u{{\vec U}}
\def\th{\vec{\theta}}
\def\rn{r_{\nu}}
\def\rnp{r'_{\nu}}

\def\HI{H~{\sc i}\,}
\def\HII{H~{\sc ii}\,}
\def\np{(\dot{N}_{phs}/10^{57}\, {\rm sec^{-1}})}
\def\tq{(\tau_Q/10^7 \, {\rm yr})}
\begin{document}
\title[Constraining QSO and IGM Properties by Bubble
  Detection]{Constraining Quasar and IGM Properties Through Bubble
  Detection in Redshifted 21-cm Maps}

\author[Majumdar, Bharadwaj \& Choudhury ]{Suman
  Majumdar$^1$\thanks{E-mail: sumanm@phy.iitkgp.ernet.in}, Somnath
  Bharadwaj$^1$\thanks{E-mail: somnath@phy.iitkgp.ernet.in} and T. Roy
  Choudhury$^{2,3}$\thanks{E-mail: tirth@hri.res.in} \\ $^1$Department
  of Physics and Meteorology \& Centre for Theoretical Studies, IIT,
  Kharagpur 721302, India\\ $^2$Harish-Chandra Research Institute,
  Chhatnag Road, Jhusi, Allahabad 211019, India\\ $^3$National Centre
  for Radio Astrophysics, TIFR, Post Bag 3, Ganeshkhind, Pune 411007,
  India}

\maketitle

\begin{abstract}
The infrared detection of a $z>7$ quasar \citep{mortlock11} has opened
up a window to directly probe the inter-galactic medium (IGM) during
the epoch of reionization. It is anticipated that future
observations will yield more quasars extending to higher redshifts. In
this paper we theoretically consider the possibility of detecting the
ionized bubble around a $z=8$ quasar using targeted redshifted 21-cm
observations with the GMRT.  The apparent shape and size of the
ionized bubble, as seen by a distant observer, depends on the
parameters $\dot{N}_{phs}/C$, $\xh1/C$ and $\tau_Q$ where
$\dot{N}_{phs}$ and $\tau_Q$ are respectively the ionizing photon
emission rate and age of the quasar, and $\xh1$ and $C$ are
respectively the neutral fraction and clumping factor of the IGM. The
21-cm detection of an ionized bubble, thus, holds the promise of
allowing us to probe the quasar and IGM properties at $z=8$.

In the current work we have analytically calculated the apparent shape
and size of a quasar's ionized bubble assuming an uniform IGM and
ignoring other ionizing sources besides the quasar, and used this as a
template for matched filter bubble search with the GMRT visibility
data.  We have assumed that $\dot{N}_{phs}$ is known from the observed
infrared spectrum, and $C=30$ from theoretical considerations, which
gives us the two free parameters $\xh1$ and $\tau_Q$ for bubble
detection. Considering $1,000$ hr of observation, we find that there
is a reasonably large region of parameter space bounded within
$(\xh1,\,\tq) = (1.0,\,0.5)$ and $(0.2,\,7.0)$ where a $3\sigma$
detection is possible if $\np=3$. The available region increases if
$\dot{N}_{phs}$ is larger, whereas we need $\xh1 \ge 0.4$ and $\tq \ge
2.0$ if $\np=1.3$. Considering parameter estimation, we find that in
many cases it will be possible to quite accurately constrain $\tau_Q$
and place a lower limit on $\xh1$ with $1,000$ hr of observation,
particularly if the bubble is in the early stage of growth and we have
a very luminous quasar or a high neutral fraction. Deeper follow up
observations ($4,000$ and $9,000$ hr) can be used to further tighten
the constraints on $\tau_Q$ and $\xh1$. We find that the
  estimated $\xh1$ is affected by uncertainty in the assumed value of
  $C$. The quasar's age $\tau_Q$ however is robust and is unaffected
  by the uncertainty in $C$.

The presence of other ionizing sources and inhomogeneities in the IGM
distort the shape and size of the quasar's ionized bubble. This is a
potential impediment for bubble detection and parameter estimation.
We have used the semi-numerical technique to simulate the apparent
shape and size of quasar ionized bubbles incorporating these
effects. If we consider a $9,000$ hr observation with the GMRT we
  find that the estimated parameters $\tau_Q$ and $\xh1$ are expected
  to be within the statistical uncertainties.
\end{abstract}
\section{Introduction}
The epoch of reionization is one of the least known periods in the
history of our Universe. According to the present understanding,
reionization of neutral hydrogen (\HI) is an extended process spanning
over the redshift range $6 \lesssim z \lesssim 15$ \citep{mitra}.  The
intergalactic medium (IGM) during this period is characterised by
bubbles of ionized hydrogen (\HII), centred around luminous
sources. Stars forming within early galaxies are believed to be the
major sources of ionizing photons during this era (for reviews see
\citealt{choudhury06,choudhury09a}). Quasars are expected to be very
rare during this epoch, but they are capable of generating larger \HII
bubbles around them with respect to their stellar
counterpart. Detection of these \HII bubbles will directly probe the
state of the local IGM around the ionizing sources. It will also
constrain the properties of the quasar, such as its luminosity and
age.

The highest redshift at which a quasar has been detected till date is
$z = 7.085$ \citep{mortlock11}. The Ly-$\alpha$ absorption spectrum of
this object reveals a highly ionized near zone of radius $2.1 \pm 0.1$
Mpc (physical). The observed near zone size is unexpectedly small when
compared with the other high redshift quasars observed in the redshift
range $6.0 < z < 6.4$ \citep{fan03, willott07, willott10}. The
ionizing photon emission rate estimated for this quasar is $1.3 \times
10^{57} {\rm sec^{-1}}$ assuming a power law for the unabsorbed
continuum emission blueward of Ly-$\alpha$ line. The quasar's age
($\tau_Q$) and local volume averaged neutral fraction ($\langle \xh1
\rangle_V$) can be estimated from the size of the quasar's near zone,
however the estimated $\tau_Q$ and $\langle \xh1 \rangle_V$ are
strongly correlated. For the same observation \citet{bolton11} have
constrained the $\tau_Q$ and $\langle \xh1 \rangle_V$ using simulated
Ly-$\alpha$ absorption spectra. This study shows that there are
several combinations of the $\tau_Q$ and $\langle \xh1 \rangle_V$ all
of which can reproduce the observed data. They find that the spectrum
observed by \citet{mortlock11} is consistent with $\tau_Q \sim 10^6$
yr with either $\langle \xh1 \rangle_V \sim 10^{-4} - 10^{-3}$
provided there is a proximate damped Ly-$\alpha$ absorber (DLA) or
$\langle \xh1 \rangle_V \sim 0.1$ without a proximate DLA. The same
observation is also consistent with $\tau_Q \sim 10^7$ yr and either
$\langle \xh1 \rangle_V \leq 10^{-4}$ with a proximate DLA or a fully
neutral IGM $\langle \xh1 \rangle_V \sim 1$ without a proximate
DLA. It is quite clear that observations of Ly-$\alpha$ absorption
spectra are limited in their ability to determine the age of the
quasar and the local neutral fraction. This limitation essentially
arises from two reasons. The first being that the IGM becomes nearly
opaque at a very low neutral fraction $\langle \xh1 \rangle_V \approx
10^{-4}$. It is not possible to distinguish between a neutral fraction
of $10^{-4}$ from a completely neutral medium, and the actual ionized
bubble around the quasar may extend far beyond the region inferred
from the Ly-$\alpha$ absorption spectra. The second limitation arises
from the fact that the Ly-$\alpha$ absorption spectrum is a pencil
beam observation along a line of sight (LoS) to the quasar. The
presence of a proximate DLA can completely change the interpretation
of the spectrum.

The redshifted 21-cm emission from neutral hydrogen in the epoch of
reionization is believed to be a powerful tool to detect \HII bubbles
around quasars. The intensity is directly proportional to the \HI
density, and it is in principle possible to probe the entire
ionization profile of the \HII bubble.  It may be possible to overcome
the limitations of Ly-$\alpha$ absorption spectra using redshifted
21-cm observations, and place better constraints on the quasar
parameters and the state of the IGM.  This is particularly motivated
by the fact that the presently functioning
{GMRT\footnote{http://www.gmrt.ncra.tifr.res.in}} \citep{swarup} and
LOFAR\footnote{http://www.lofar.org/}, and the upcoming
{MWA\footnote{http://www.haystack.mit.edu/ast/arrays/mwa/}} and {21CMA
\footnote{http://21cma.bao.ac.cn/}} are all sensitive to the \HI
signal from the epoch or reionization. However, redshifted \HI
observations have their own limitations in that the \HI signal is
extremely small relative to the sensitivities of the present and
upcoming telescopes. Further, the signal will be buried deep in
foregrounds which are a few orders of magnitude larger than the signal
\citep{ali08,bernardi09}.

\citet{datta2} (hitherto Paper I) have proposed a matched filter
technique to detect ionized bubbles in radio-interferometric
observations of redshifted 21-cm emission. The matched filter
optimally combines the entire signal of an ionized bubble while
minimizing the noise and the foreground contributions. The redshift $z
\sim 8$ is most optimal for bubble detection \citep{datta09}, and at
this redshift a $3\sigma$ detection is possible for bubbles of
comoving radius larger than $24$ and $28$ Mpc with $1000$ hr of
observation using the GMRT and MWA respectively, assuming that the
neutral hydrogen fraction is $\sim 0.6$ outside the bubble. This
technique, however, is limited by the density fluctuations in \HI
outside the bubble, and analytical estimates (Paper I) and simulations
\citep{datta3} show that it is not possible to detect bubbles of
comoving radius $\leq 6$ and $\leq 12$ Mpc using the GMRT and MWA
respectively, however large be the observation time.

The matched filter technique mentioned above assumes the \HII bubble
to be spherical. However, a growing spherical bubble will appear
anisotropic along the LoS to a present day observer due to the finite
light travel time (FLTT) \citep{wyithe04,yu05,wyithe05,sethi08} and
also due to the evolution of global ionization fraction
\citep{geil2}. In an earlier work (\citet{majumdar11}; hitherto Paper
II), we have analytically quantified this anisotropy and studied the
possibility of detecting such a bubble in a targeted search around a
known quasar.  We find that the bubble appears elongated along the LoS
during the early stages of its growth, whereas it appears compressed
in the later stages when the growth slows down.  In addition to this,
the apparent centre of the bubble also shifts towards the observer. We
find that a spherical filter is adequate for bubble detection even
when the apparent anisotropy of the bubble's shape is taken into
account, the centre of the best matched filter will however be shifted
relative to the quasar. We also propose that the measured shift and
the radius of the best matched filter can together be used to
constrain the age and luminosity of the quasar.
  
In this work we consider the detection of an \HII bubble in a targeted
21-cm search centred on a known quasar at $z = 8$ and we investigate
the possibility of using such an observation to constrain the
properties of the quasar and the IGM.  For this purpose we improve
upon the spherical filter by calculating the apparent, anisotropic
shape of the quasar bubble and using this as a template for the
filter.We expect this to give a better match to the bubble that is
actually present in the data. Further, we also expect improved
parameter estimation using the improved filter.

Our earlier work (Paper II) was entirely based on analytic estimates
which do not take into account the presence of ionizing sources other
than the quasar and the inhomogeneities in the IGM.  We have overcome
this limitation here by using the semi-numerical technique
\citep{choudhury09b} to simulate the ionization field.

This paper is arranged as follows -- In Section 2 we briefly discuss
the model for bubble growth around a quasar, and the matched filter
technique for detecting such a bubble. We also present the improved
filter based on the calculated apparent shape of the bubble. In
Section 3 we use analytic estimates, based on the bubble growth
equation, to study the SNR for bubble detection. We use this to
determine the parameter range where a $3\sigma$ detection is possible
in $1,000$ hr of observation with the GMRT. We then consider parameter
estimation, and explore the kind of limits that can be placed on the
quasar and IGM properties using the matched filter technique. Section
4 briefly describes how we have simulated the apparent shape of quasar
bubbles, and we present our results and summarize our findings in
Section 5.

Our entire analysis is restricted to the  redshift $z=8$. We have used 
the values from the WMAP 7 year data \citep{komatsu,jarosik} $h= 0.705
$, $\Omega_m = 0.2726$, $\Omega_{\Lambda} = 0.726$, $\Omega_b h^2 = 
0.0223$ for the cosmological parameters. 

\section{Bubble Detection}
\subsection{Model for Bubble Growth}
\begin{figure}
\psfrag{C}[c][c][1][0]{{\bf {\large C}}}
\psfrag{Q}[c][c][1][0]{{\bf {\large Q}}}
\psfrag{A}[c][c][1][0]{{\bf {\large A}}}
\psfrag{B}[c][c][1][0]{{\bf {\large B}}}
\psfrag{D}[c][c][1][0]{{\bf {\large D}}}
\psfrag{E}[c][c][1][0]{{\bf {\large E}}}
\psfrag{LOS}[c][c][1][0]{{\bf {\large LoS}}}
\psfrag{P}[c][c][1][0]{{\bf {\Large ${\boldsymbol \phi}$}}}
\includegraphics[width=.45\textwidth, angle=0]{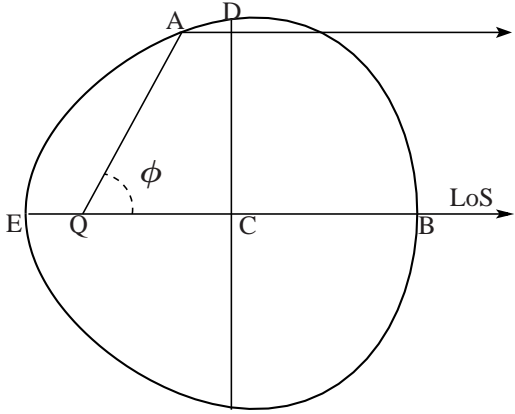}
\caption{This shows the apparent shape of the \HII bubble around a
  quasar Q as seen by a distant observer located in the direction
  indicated by the arrows. The apparent anisotropy of the \HII bubble
  arises from the fact that photons received from different parts of
  the bubble had been emitted at different stages of of its growth.
  The apparent centre of the bubble is shifted by $QC$ from the quasar
  Q to the point C. The bubble's apparent radius is $EC = CB =
  R_{\parallel}$ along the LoS and $CD = R_{\perp}$ perpendicular to
  LoS. }
\label{fig:shape}
\end{figure}
The growth of a spherical \HII region around a quasar which is
isotropically emitting ionizing photons at a rate $\dot{N}_{phs}$ is
governed by the equation
\citep{shapiro87,white03,wyithe04,wyithe05,yu05,sethi08},
\begin{equation}
\frac{4\pi}{3} \frac{d}{d\tau}\left(x_{HI} \langle n_H \rangle
r^3\right)=\dot{N}_{phs} -\frac{4}{3}\pi \alpha_B C \langle n_H
\rangle^2 r^3 \,,
\label{eq:yu}
\end{equation}
[eq. (1) of Paper II and eq (7) of \citealt{yu05}]. It is assumed that
the quasar is triggered at a cosmic time $t_i$, and $\tau = t - t_i$
denotes the quasar's age at any later time $t$. The variable $r(\tau)$
denotes the radius of the spherical ionizing front at the instant when
a photon that was emitted from the quasar at $\tau$ catches up with
the ionizing front. The term $\alpha_B(=2.6\times 10^{-13} \, {\rm
  cm}^3 \, s^{-1})$ is the recombination coefficient to excited levels
of hydrogen at $T=10^4 \, {\rm K}$, $\langle n_H \rangle$ and $\langle
n_{\HI} \rangle$ are the average hydrogen and neutral hydrogen
densities respectively, $\xh1$ is the neutral fraction of the IGM and
$C \equiv \langle n^2_{\HI} \rangle/\langle n_H \rangle^2$ is the
clumping factor which quantifies the effective clumpiness of the
hydrogen inside the bubble. Eq. (\ref{eq:yu}) tells us that the growth
of the \HII bubble is determined by the ionizing photon emission rate
after accounting for the photons required to compensate for the
recombinations inside the existing ionized region.

For a constant $\dot{N}_{phs}$ the solution to the above growth model
(eq. (\ref{eq:yu})) takes the form
\begin{equation}  
r(\tau)=r_S
\left[1-\exp\left(-\frac{\tau}{\tau_{rec}}\right)
  \right]^{\frac{1}{3}}\,,
\label{eq:growth}
\end{equation} 
[eq. (2) of Paper II and eq. (8) of \citealt{yu05}] where $\tau_{rec} = \xh1
\left(C\,\langle n_H \rangle\,\alpha_B\right)^{-1}$ and
$r_S = \left(3\dot{N}_{phs} \tau_{rec}/(4 \pi \xh1 \langle n_H \rangle
) \right)^{1/3}$.
The apparent shape of the growing bubble that is perceived by a present
day observer will be distorted due to the fact that photons received
by the observer from different parts of the bubble had been emitted at
different stages of the bubble's growth (see Figure
\ref{fig:shape}). To visualize this apparent shape one need the
relation between $r$ and the angle $\phi$ between the LoS and the
point A (Figure \ref{fig:shape}) under consideration on the ionization
front. Consider a quasar observed at an age $\tau_Q$.  The light travel time
starting from the quasar at $\tau$ to the point A and then to the
observer is $[r(\tau)/c](1-\cos \phi)$ more compared to the photon
that was emitted from the quasar at $\tau_Q$ and travels
straight to the observer. This gives
\begin{equation} 
\tau_Q=\tau +\frac{r(\tau)}{c}(1-\cos\,\phi)\,,
\label{eq:shape}
\end{equation} 
[eq. (7) of Paper II and eq. (3) of \citealt{yu05}]. We use 
eq. (\ref{eq:growth}) and eq. (\ref{eq:shape}) to determine $r$ as a
function of $\phi$. This gives the quasar's apparent shape shown in
Figure \ref{fig:shape}. 

The apparent shape of a quasar's \HII bubble has been studied in
detail in Paper II. We find that the FLTT has two effects, 1. the
bubble appears to be anisotropic along the LoS and 2. the bubble's
apparent centre shifts along the LoS away from the quasar towards the
observer. We now briefly discuss how we parametrize these two
effects. Referring to Fig \ref{fig:shape}, we use $R_{\perp} = {\rm
  CD}\,$ to quantify the overall comoving size of the bubble. The
bubble centre C is located mid way between EB. We use the
dimensionless shift parameter $s$ defined as,
\begin{equation}
s = \frac{{\rm QC}}{R_{\perp}} \,,
\end{equation}
to quantify the shift in the bubble's centre as a fraction of its
radius. We find (see Paper II for details) that $s$ can be greater
than $1$ during the early stages of the bubble's growth and it
approaches $0$ in the later stages of evolution.

We use the dimensionless anisotropy parameter $\eta$ defined as,
\begin{equation}
\eta =\frac{{\rm CB}}{R_{\perp}} - 1 \,,
\end{equation}
to quantify the anisotropy in the apparent shape of the bubble. A
value $\eta > 0$ indicates that the bubble is elongated along the LoS
and a value $\eta <0$, indicates that it is compressed along the
LoS. Our earlier work (Paper II) shows that $\eta$ has values in the
range $0.1$ to $0.5$ during the early stages of  growth when the 
bubble appears elongated along the LoS. We also find that $\eta$ has
values in the range $0.0$ to $-0.2$ in the later stages of
evolution when the bubble may appear compressed along the LoS. 

\subsection{Matched filter bubble detection}
\label{subsec:ani_filt}
The basic observable quantity in the radio interferometry is the 
visibility ${\mathcal V}(\u,\nu)$ which is related to the specific intensity
pattern on the sky $I_{\nu}(\th)$ as
\begin{equation}
{\mathcal V}(\u,\nu)=\int d^2 \theta A(\th) I_{\nu}(\th)
e^{ 2\pi \imath \th \cdot \u}
\label{eq:vis}
\end{equation}
The baseline $\u ={\vec d}/\lambda$ denotes the antenna separation
${\vec d}$ projected in the plane perpendicular to the LoS in units of
the observing wavelength $\lambda$, $\th$ is a two dimensional vector
in the sky plane with the origin at the centre of the FoV (the
phase centre) and $A(\th)$ is the beam pattern of a single
antenna. For the GMRT $A(\th)$ can be well approximated by a
Gaussian $A(\th)=e^{-{\theta}^2/{\theta_0}^2}$ where $\theta_0 \approx
0.6 ~\theta_{\rm FWHM}$ and we use $2.28^{\circ}$ for $\theta_0$ at
$157.77$ MHz {\it i.e.} $z = 8$. We consider a situation where the
observation is spanned across several frequency channels around the
central frequency of $157.77$ MHz.

The visibility recorded in a radio-interferometric observation is
actually a combination of several contributions, 
\begin{equation}
{\mathcal V}(\vec{U},\nu)=S(\vec{U},\nu)+HF(\vec{U},\nu)+N(\vec{U},\nu)+F(\vec{U},\nu)\,,
\label{eq:vis_comp}
\end{equation}
where the first term $S(\vec{U},\nu)$ is the signal from the ionized
region that is actually present in the observational data,
$HF(\vec{U},\nu)$ is the contribution from fluctuations in the \HI
distribution outside the ionized bubble, $N(\vec{U},\nu) $ is the
system noise which is inherent to the measurement and $F(\vec{U},\nu)$
is the contribution from astrophysical foregrounds. 

The signal $S(\vec{U},\nu)$ from an ionized region will appear as a
decrement with respect to the background 21-cm radiation. A spherical
\HII bubble embedded in an uniform IGM with a neutral fraction $\xh1$
is parametrized by its comoving radius $R_b$, the redshift $z_c$ and
the angular position $\th_c$ corresponding to the centre of the
bubble. The quantity $\rn$ is the comoving distance to the redshift
where the \HI emission received at $\nu=1420 \, {\rm MHz}/(1+z)$
originated, and $\rnp=d \, \rn/d \, \nu$. A plane perpendicular to LoS
and passing through the centre of the bubble at a comoving distance
$\rn$ will cut a disk of comoving radius $R_{\nu}=R_b \sqrt{1- (\Delta
  \nu/\Delta \nu_b)^2}$ from the bubble, where $\Delta \nu=\nu_c-\nu$
and $\Delta \nu_b=R_b/r'_{\nu_c}$, and $\theta_{\nu}=R_{\nu}/r_{\nu}$
is the angular extent corresponding to $R_{\nu}$. For a bubble located
at the centre of the FoV we have,
\begin{align}
S(\u,\nu) = -\pi \bar{I_{\nu}} \xh1 \theta^2_\nu 
&\left [ \frac{2 J_1(2 \pi U \theta_\nu
  )}{2 \pi U \theta_\nu}\right ] \nonumber \\
&\Theta \left(1- \frac{\mid \nu -\nu_c \mid}{\Delta \nu_b} \right) \,,
\label{eq:signal}
\end{align}
where $\bar{I}_{\nu}=2.5\times10^2\frac{Jy}{sr} \left (\frac{\Omega_b
  h^2}{0.02}\right )\left( \frac{0.7}{h} \right ) \left
(\frac{H_0}{H(z)} \right )$,  $\Theta(x)$ is the Heaviside step
function, and $J_1(x)$ is the Bessel function of first order. The
expected signal has  a  peak value (Paper I) of  $S(0,\nu)=1.12\, {\rm
  mJy}$ for  a bubble with  $R_b = 40$ Mpc, and $S(0,\nu) \propto R_b^2$.

The expected \HI signal is very weak, so much so that the contribution
$HF(\vec{U},\nu)$ from \HI fluctuations outside the bubble may, in
some cases, exceed the expected signal. We consider the system noise
$N(\vec{U},\nu)$ in each baseline and frequency channel
to be an independent Gaussian random variable with zero
mean and r.m.s.
\begin{equation}
\sqrt{\langle \hat{N}^2 \rangle} = C^x \left( \frac{\Delta \nu}{1 {\rm
 MHz}} \right)^{-1/2} \left( \frac{\Delta t}{1 sec}\right)^{-1/2}
\label{eq:noise}
\end{equation}
where the constant $C^x$ is different for different interferometric
arrays, $\Delta \nu$ is the channel width and $\Delta t$ is the
correlator integration time. Following Paper I, we use $C^x = 1.03$ Jy
for the GMRT. We note that the r.m.s. noise $\sqrt{\langle \hat{N}^2
  \rangle}$ will actually vary with $\vec{U}$, we have ignored this
baseline dependence in order to keep the analysis simple. Further we
do not expect this to be a very significant effect compared to the
various other uncertainties arising from the lack of knowledge about
the IGM and the quasar properties. The astrophysical foregrounds
$F(\vec{U},\nu)$ are expected to be several orders of magnitude
larger than the \HI signal. But they are predicted to have a
featureless, continuum spectra whereas the signal is expected to have
a dip around $\nu_c$ (central frequency of the target \HII
bubble). This difference holds the promise of allowing us to separate
the signal from the foregrounds.

The visibility based matched filter technique, proposed in Paper I,
optimally combines the signal from an ionized region while minimizing
the other contributions. In this technique we use the signal expected
from a spherical \HII bubble centered at redshift $z_b$ and comoving
radius $R_b$ as a template, and search for the presence of this signal
in the data using the estimator,
\begin{equation} 
\hat{E}(z_b, R_b )= \sum_{a,b} S_{f}^{\ast}(\u_a,\nu_b) \hat{{\mathcal
    V}}(\u_a,\nu_b)\,. 
\label{eq:estim0}
\end{equation}
Here $S_{f}(\u,\nu)$ is the filter and $\hat{{\mathcal
    V}}(\u,\nu)$ are the measured visibilities. The filter is defined
as,
\begin{align}
S_{f}(\u,\nu) =& \left(\frac{\nu}{\nu_c}\right)^2 \left[S(\u,\nu)
  - \right.\nonumber\\ 
&\left. \frac{\Theta(1 - 2|\nu - \nu_c|/B^{\prime})}{B^{\prime}} \int^{\nu_c
  +B^{\prime}/2}_{\nu_c - B^{\prime}/2} S(\u,\nu^{\prime})
  d\nu^{\prime} \right]\,,
\label{eq:filter}
\end{align}
where $S(\u,\nu)$ is the signal expected from the bubble that we
  are trying to detect. The filter eliminates any frequency
  independent component of the foreground from the frequency range
  $\nu_c + B^{\prime}/2$ to $\nu_c - B^{\prime}/2$. All the
contributions to ${\mathcal V}$, except the signal $S(\u,\nu)$, are
assumed to be random variables of zero mean, uncorrelated to the
filter whereby the estimator has expectation value,
\begin{equation}
\langle \hat{E} \rangle=\sum_{a,b} S_{f}^{\ast}(\u_a,\nu_b) \,
S(\u_a,\nu_b)\, .  
\end{equation}
The other terms in eq. (\ref{eq:vis_comp}) contribute only to the
variance,
\begin{equation} 
\langle (\Delta \hat{E})^2 \rangle
= \langle (\Delta \hat{E})^2 \rangle_{{\rm HF}} +\langle (\Delta
\hat{E})^2\rangle_{{\rm N}} + \langle (\Delta \hat{E})^2\rangle_{{\rm
    F}}\,.  
\end{equation}
The signal to noise ratio for the estimator is
\begin{equation}
{\rm SNR}=\frac{\langle \hat{E} \rangle}{\sqrt{\langle (\Delta
    \hat{E})^2 \rangle}}
\end{equation}
Matched filter bubble detection is carried out by analyzing the
SNR for different values of the filter parameters $z_f$ and $R_f$. We
expect the SNR to peak when the filter parameters $z_f$ and $R_f$
exactly match the parameters $z_b$ and $R_b$ of the bubble that is
actually present in the data. We have a statistically significant
($3\sigma$) bubble detection if the peak ${\rm SNR} \ge 3$. Note that
the two angular coordinates $\theta_x$ and $\theta_y$ are already
known in a targeted search around a quasar. In general, $\theta_x$ and
$\theta_y$ have also to be treated as free parameters in a blind search. 
 
The filter $S_f(\u,\nu)$ is defined in such a way
[eq. (\ref{eq:filter})] that it is insensitive to the presence of a
smooth frequency independent foreground component. Under the assumed
foreground model (Paper I), the residual foreground contribution
$\langle (\Delta \hat{E})^2\rangle_{{\rm F}}$ is predicted to be much
smaller than the other contributions to $\langle (\Delta \hat{E})^2
\rangle$ and we do not consider it here. In the absence of patchy
reionization outside the bubble, the \HI fluctuations trace the dark
matter fluctuations. This imposes a lower bound $R_b > 12$ Mpc for the
GMRT on the smallest \HII bubble that can be detected
\citep{datta3}. Considering patchy reionization outside the bubble, we
expect galaxies to produce ionized regions with a typical radius of $6
\, {\rm Mpc}$ or smaller.  The quasar bubbles that we can detect using
the GMRT $(\ge 20 \, {\rm Mpc})$ are much larger than these ionized
patches.  Earlier work \citep{datta3} shows that the \HI fluctuations
do not play a very important role if the quasar bubble is much larger
than the typical size of the ionized patches outside the bubble, and
we have ignored this contribution in our estimates in this paper.  We
have only considered the system noise $\langle (\Delta
\hat{E})^2\rangle_{{\rm N}}$ which is the most dominant component in
$\langle (\Delta \hat{E})^2 \rangle$. The resulting SNR is inversely
proportional to the square root of the total observation time SNR
$\propto t_{obs}^{1/2}$.
\begin{figure*}
\psfrag{tq=0.5}[c][c][1][0]{{\bf {\large $\tau_Q / 10^7$ yr$=0.5$}}}
\psfrag{tq=2.0}[c][c][1][0]{{\bf {\large $\tau_Q / 10^7$ yr$=2.0$}}}
\psfrag{tq=4.0}[c][c][1][0]{{\bf {\large $\tau_Q / 10^7$ yr$=4.0$}}}
\psfrag{Mpc}[c][c][1][0]{{\bf {\Large Mpc}}}
\includegraphics[width=.4\textwidth,
  angle=-90]{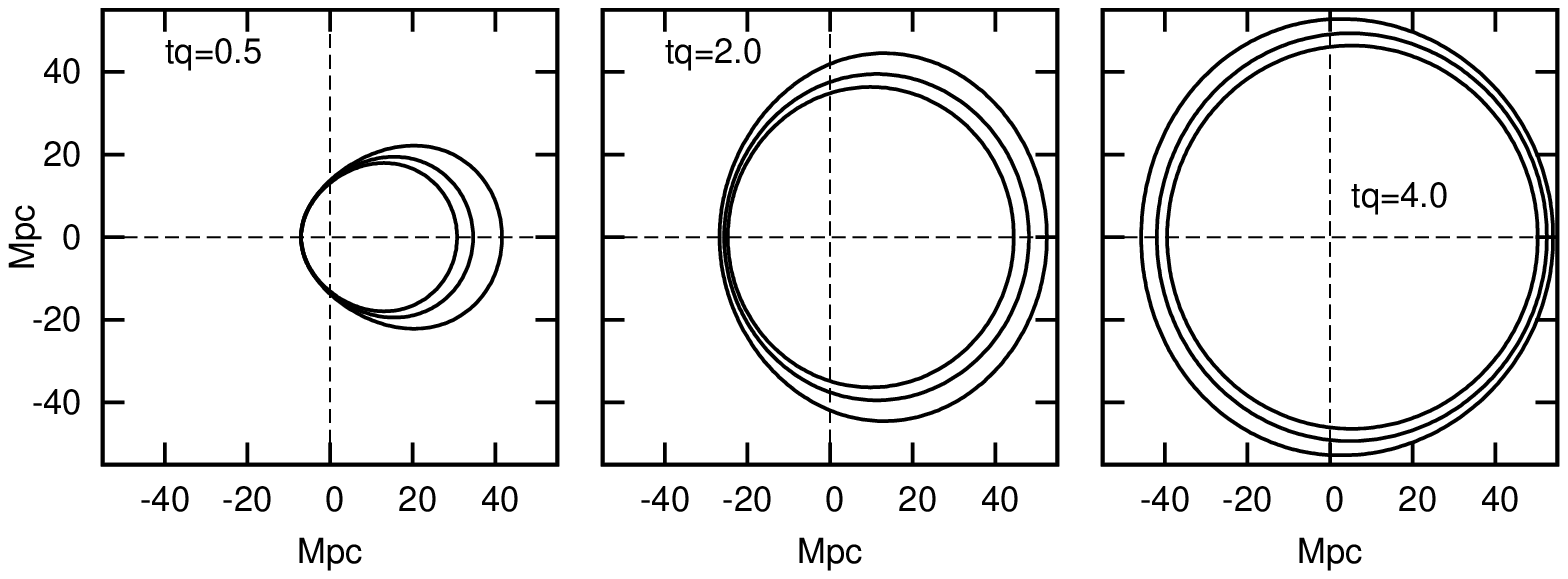}
\caption{This shows the \HII bubble around a quasar with $\np=3$ at
  different stages of growth ($\tq = 0.5,\, 2.0,\, 4.0$). The three
  different shapes in each panel correspond to different IGM neutral
  fractions ($\xh1 = 0.75,\, 0.50,\, 0.25$), the bubble size
  increasing with decreasing neutral fraction.  Note that ($\tq =
  2.0$) corresponds to $\tau_Q/\tau_{rec}=1.21$ for $\xh1=0.5$ and
  $\tau_{rec} \propto \xh1$.  The quasar is at the centre of each
  panel. The x-axis represents the LoS and the observer is on the
  right side of the panels.}
\label{fig:shape_xh1}
\end{figure*}
\begin{table}
\centering
\begin{tabular}{c|c|c|c|c}
\hline
\hline
 $\tau_Q/10^7$ & $\xh1$ & $R_{\perp}$ & $\eta$ & $s$ \\
 (yr) &  & (Mpc) &  &  \\
\hline
 $0.5$ & $0.75$ & $17.9$ & $0.05$ & $0.66$ \\
 $0.5$ & $0.50$ & $19.4$ & $0.07$ & $0.71$ \\
 $0.5$ & $0.25$ & $22.0$ & $0.11$ & $0.79$ \\
 $2.0$ & $0.75$ & $36.3$ & $-0.05$ & $0.27$ \\
 $2.0$ & $0.50$ & $39.4$ & $-0.07$ & $0.29$ \\
 $2.0$ & $0.25$ & $44.5$ & $-0.11$ & $0.29$ \\
 $4.0$ & $0.75$ & $46.3$ & $-0.03$ & $0.12$ \\
 $4.0$ & $0.50$ & $49.3$ & $-0.04$ & $0.11$ \\
 $4.0$ & $0.25$ & $52.8$ & $-0.06$ & $0.07$ \\
\hline
\hline
\end{tabular}
\caption{This tabulates the size ($R_{\perp}$), and  the anisotropy
  and shift parameters ($\eta$   and $s$ respectively)  corresponding
  to the shapes shown in   Figure \ref{fig:shape_xh1}.}
\label{tab:eta_s}
\end{table}
\subsection{Anisotropic filter} 
The matched filter technique considered in earlier works [Paper I;
  \citet{datta3,datta09}] and briefly described above uses the signal
expected from a spherical \HII bubble [eq. (\ref{eq:signal})] as a
template for the filter [eq. (\ref{eq:filter})]. The actual bubble is
however anisotropic due to the FLTT, and we expect a mismatch between
the template and the actual signal. We can avoid this if we use the
apparent anisotropic shape as the template for the filter $S_f(\u,
\nu)$. This is done by numerically determining the bubble radius $r$
as a function of $\phi$ (Figure \ref{fig:shape}) by solving
eq. (\ref{eq:growth}) and (\ref{eq:shape}). A section through the
anisotropic bubble continues to be a circular disk, and
eq. (\ref{eq:signal}) still holds for the signal. The only difference
is that the comoving radius $R_b$ now varies along the LoS, and we
have to use the calculated $r(\phi)$ instead of a fixed $R_b$. An
inspection of eq.s (\ref{eq:yu}), (\ref{eq:growth}) and
(\ref{eq:shape}) shows that the apparent shape is completely specified
by three free parameters $\dot{N}_{phs}/C$, $\xh1/C$ and $\tau_Q$. It
is therefore necessary to analyze the SNR for different values of
these three parameters and determine the values of $\dot{N}_{phs}/C$,
$\xh1/C$ and $\tau_Q$ for which the SNR peaks.

Here we consider a targeted search around a known quasar whose
infrared spectrum has been measured. It is possible to extrapolate the
observed quasar's infrared spectrum blueward of the Ly-$\alpha$ line
and estimate $\dot{N}_{phs}$ (example \citealt{mortlock11}). It is
reasonable to assume that $\dot{N}_{phs}$ is known in a targeted search
for bubble detection. Further, the analysis of numerical simulations
indicate that the clumping factor has a value $C=30$ at $z=8$
\citep{gnedin97,yu05,yu05b}. With this assumption the apparent shape
of the \HII bubble is completely specified by only two parameters the
mean neutral fraction $\xh1$ and $\tau_Q$ the age of the quasar. Our
entire analysis of bubble detection is in terms of these two
parameters, $\xh1$ and $\tau_Q$. For a quasar with a constant
luminosity $\np=3$, Figure \ref{fig:shape_xh1} shows how the shape and
size of the \HII bubble vary with $\xh1$ and $\tau_Q$.  The bubble
grows as $r(\tau) \propto (\tau/\xh1)^{1/3}$ in the early stages when
$\tau \ll \tau_{rec}$ (eq. \ref{eq:growth}) and saturates at
$r(\tau)=r_s$, which is independent of $\xh1$, when
$\tau/\tau_{rec}\sim 1$. Figure \ref{fig:shape_xh1} and
Table~\ref{tab:eta_s} show that the size of the bubble decreases with
increasing $\xh1$ during the early stage of growth, and this effect
becomes less noticeable at the later stage of growth. The bubble
appears elongated along the LoS in the early stage, whereas it becomes
compressed in the later stage.  We see that the size $R_{\perp}$,
anisotropy $\eta$ and shift $s$ all vary with $\tau_Q$ and $\xh1$.

\section{Analytic Estimates}
\begin{figure*}
\psfrag{xh1}[c][c][1][0]{{\bf {\Large $\xh1$}}}
\psfrag{tq}[c][c][1][0]{{\bf {\Large $\tau_Q / 10^7\,$ yr}}}
\psfrag{1000hr}[c][c][1][0]{{\textcolor{white}{{\bf {\large $1000$ hr}}}}}
\psfrag{2000hr}[c][c][1][0]{{\textcolor{white}{{\bf {\large $2000$ hr}}}}}
\psfrag{4000hr}[c][c][1][0]{{\bf {\large $4000$ hr}}}
\psfrag{8000hr}[c][c][1][0]{{\bf {\large $8000$ hr}}}
\psfrag{3sigma}[c][c][1][0]{{\bf {\huge $3\sigma$}}}
\psfrag{5sigma}[c][c][1][0]{{\bf {\huge $5\sigma$}}}
\includegraphics[width=.4\textwidth,
  angle=-90]{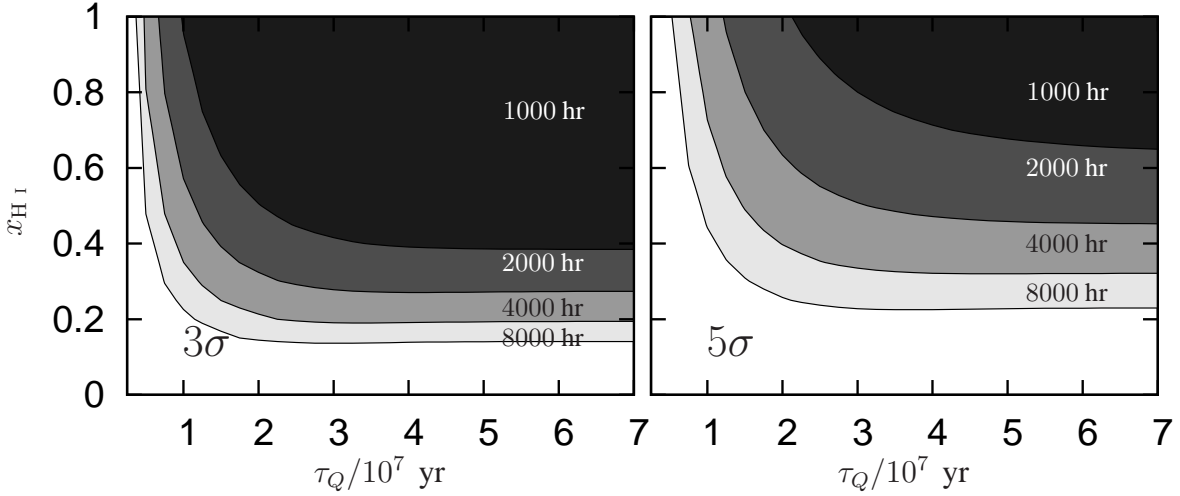}
\caption{This shows the analytic estimates of minimum observation time required
  for $3\sigma$ and $5\sigma$ (left and right panels respectively)
  detection of the \HII bubble around the quasar ULASJ1120+0641
  discovered by \citet{mortlock11}.}
\label{fig:mort}
\end{figure*}
\begin{figure*}
\psfrag{xh1}[c][c][1][0]{{\bf {\LARGE $\xh1$}}}
\psfrag{tq}[c][c][1][0]{{\bf {\LARGE $\tau_Q / 10^7\,$ yr}}}
\psfrag{20}[c][c][1][0]{{\Large $20$}}
\psfrag{30}[c][c][1][0]{{\Large $30$}}
\psfrag{40}[c][c][1][0]{{\Large $40$}}
\psfrag{50}[c][c][1][0]{{\Large $50$}}
\psfrag{54}[c][c][1][0]{{\Large $54$}}
\psfrag{60}[c][c][1][0]{{\Large $60$}}
\psfrag{70}[c][c][1][0]{{\Large $70$}}
\psfrag{75}[c][c][1][0]{{\Large $75$}}
\psfrag{1s}[c][c][1][0]{{\Large $1\sigma$}}
\psfrag{3s}[c][c][1][0]{{\Large $3\sigma$}}
\psfrag{5s}[c][c][1][0]{{\Large $5\sigma$}}
\psfrag{7s}[c][c][1][0]{{\Large $7\sigma$}}
\psfrag{9s}[c][c][1][0]{{\Large $9\sigma$}}
\psfrag{11s}[c][c][1][0]{{\Large $11\sigma$}}
\psfrag{13s}[c][c][1][0]{{\Large $13\sigma$}}
\psfrag{15s}[c][c][1][0]{{\Large $15\sigma$}}
\psfrag{17s}[c][c][1][0]{{\Large $17\sigma$}}
\psfrag{19s}[c][c][1][0]{{\Large $19\sigma$}}
\includegraphics[width=1.\textwidth, angle=0]{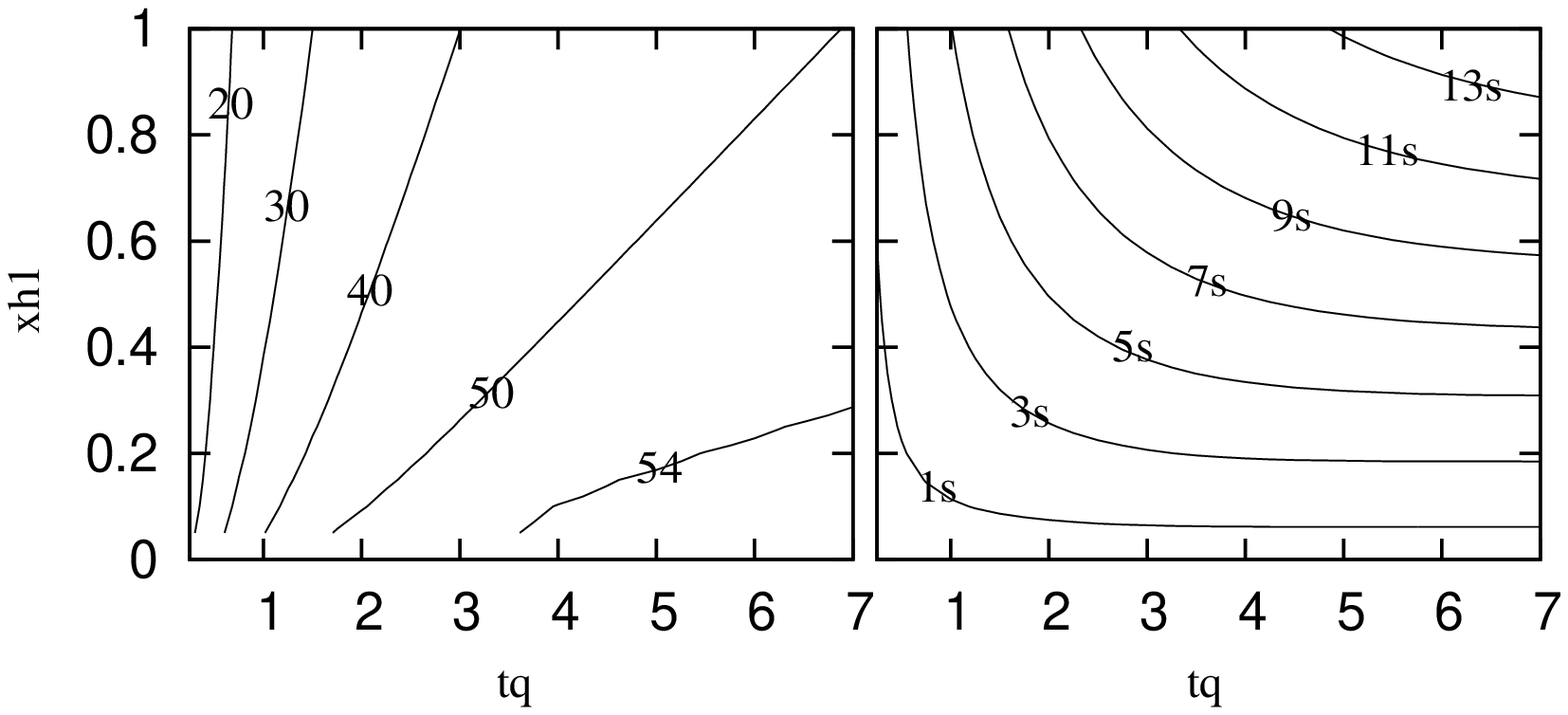}
\includegraphics[width=1.\textwidth, angle=0]{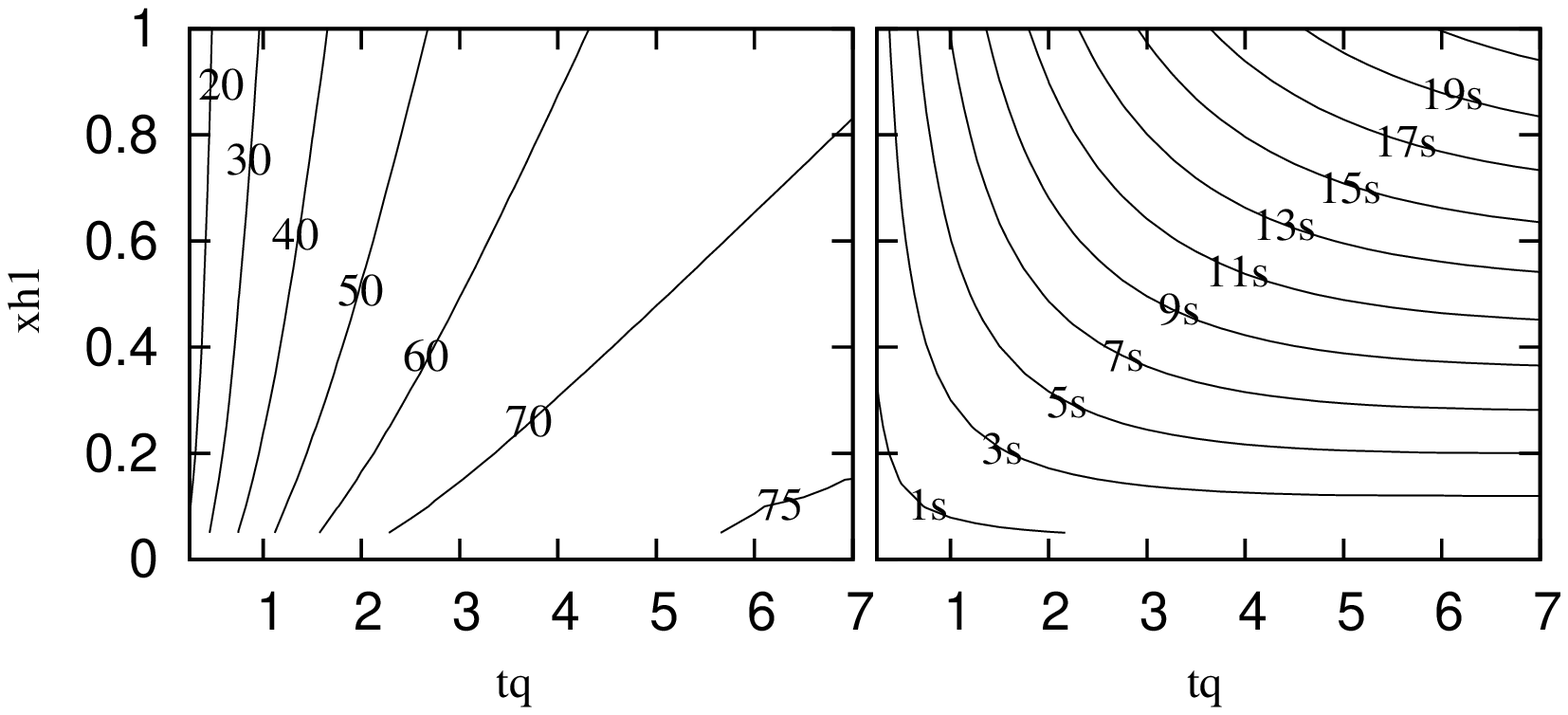}
\caption{This shows analytic estimates of the bubble's comoving  
  size ($R_{\perp}$ in Mpc)(Left panel) and the SNR for the best matched
  filter (Right panel). Top and bottom panels correspond to quasar
  photon emission rates of  $\np=3$ and $8$ respectively.}
\label{fig:contour_snr_rperp}
\end{figure*}

In this section we present analytic estimates of bubble detection in a
targeted search around a known high redshift quasar. For this purpose
the \HI outside the bubble is assumed to follow the dark matter. For
the growth of the quasar bubble we ignore the effect of other ionizing
sources and inhomogeneities in the \HI distributions. The bubble's
evolution is assumed to be exactly described by eq. (\ref{eq:growth})
discussed earlier. For bubble detection we use the apparent shape
obtained using eqs. (\ref{eq:growth}) and (\ref{eq:shape}) to calculate
the filter $S_f(\u, \nu)$. The search has two parameters $\xh1$ and
$\tau_Q$, and we expect a perfect match when these have the same value
as the bubble actually present in the data.

Quasars typically have a luminosity of around $\dot{N}_{phs} \sim
10^{57}\, {\rm sec^{-1}}$. The highest redshift quasar detected till
date has a luminosity of $\dot{N}_{phs} \sim 1.3 \times 10^{57}\, {\rm
  sec^{-1}}$ \citep{mortlock11}. Very little is known about the
luminosity distribution of very high redshift quasars. The more
luminous quasars are expected to produce larger ionized bubbles, and
the prospect of a 21-cm detection is also expected to be higher for a
more luminous quasar. In much of our analysis we have used two
different quasar luminosities $\dot{N}_{phs} = 3 \times 10^{57}\, {\rm
  sec^{-1}}$ and $8 \times 10^{57}\, {\rm sec^{-1}}$. The typical
quasar age is expected to be in the range $10^6 - 10^8$ yrs
\citep{haehnelt98,haiman01}. The analysis of the proximity zones in
the Ly-$\alpha$ spectra of high redshift quasars yield quasar ages
which are larger than $1 - 3 \times 10^7$ yrs
\citep{worseck06,lu11}. We have considered quasar ages in the range
$10^6 - 10^8$ yrs in our analysis.
\begin{figure*}
\psfrag{xh1}[c][c][1][0]{{\bf {\LARGE $\xh1$}}}
\psfrag{tq}[c][c][1][0]{{\bf {\LARGE $\tau_Q / 10^7\,$ yr}}}
\psfrag{0.10}[c][c][1][0]{{\Large $0.10$}}
\psfrag{0.05}[c][c][1][0]{{\Large $0.05$}}
\psfrag{0.00}[c][c][1][0]{{\Large $0.00$}}
\psfrag{-0.05}[c][c][1][0]{{\Large $-0.05$}}
\psfrag{-0.10}[c][c][1][0]{{\Large $-0.10$}}
\psfrag{-0.15}[c][c][1][0]{{\Large $-0.15$}}
\psfrag{-0.20}[c][c][1][0]{{\Large $-0.20$}}
\psfrag{0.1}[c][c][1][0]{{\Large $0.1$}}
\psfrag{0.2}[c][c][1][0]{{\Large $0.2$}}
\psfrag{0.3}[c][c][1][0]{{\Large $0.3$}}
\psfrag{0.4}[c][c][1][0]{{\Large $0.4$}}
\psfrag{0.5}[c][c][1][0]{{\Large $0.5$}}
\psfrag{0.6}[c][c][1][0]{{\Large $0.6$}}
\psfrag{0.7}[c][c][1][0]{{\Large $0.7$}}
\includegraphics[width=1.\textwidth, angle=0]{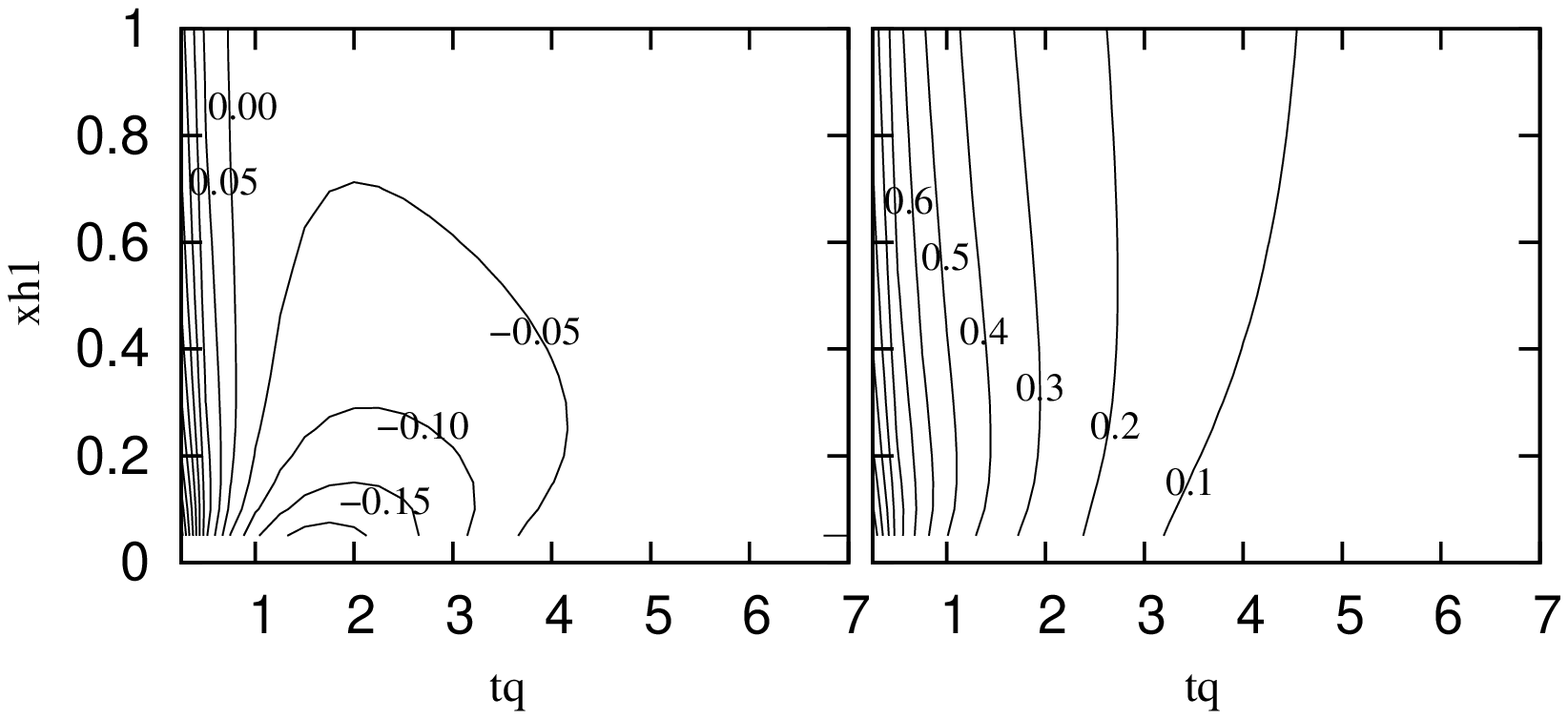}
\includegraphics[width=1.\textwidth, angle=0]{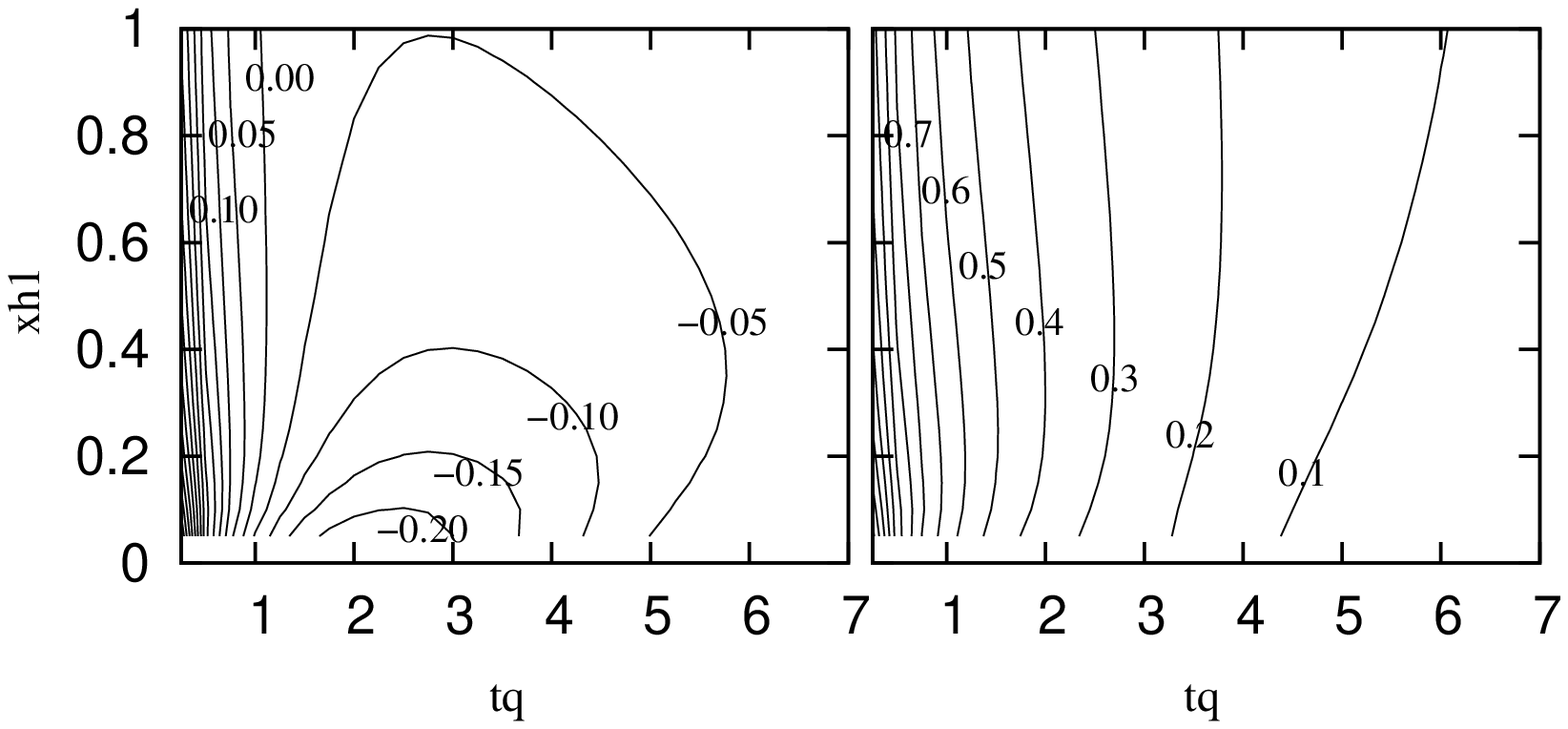}
\caption{This shows analytic estimates of the anisotropy 
  parameter $\eta$ (Left panel) and the shift parameter $s$ (Right
  panel) for the   quasar's \HII bubble. Top and bottom panels
  correspond to quasar photon emission rates of $\np=3$ and $8$
  respectively.} 
\label{fig:contour_eta_s}
\end{figure*}

\subsection{Detectability}

We first consider the possibility of detecting the \HII bubble around
the quasar ULASJ1120+0641 \citep{mortlock11} using GMRT redshifted
21-cm observation. The size and shape of the \HII bubble depend on
$\xh1$ and $\tau_Q$, and we have determined the observation time that
will be required for different values of these parameters. The left
and right panels of Figure \ref{fig:mort} show the observation time
that will be required for a $3\sigma$ and $5\sigma$ detection
respectively. We find that for $1,000$ hr of observation a $3\sigma$
detection is possible in a considerable region of parameter space
where $\xh1>0.4$ and $\tq>1.5$, and a $5\sigma$ detection is possible
in a relatively smaller region of parameter space where $\xh1>0.7$ and
$\tq>3.5$ . A $3\sigma$ detection is possible for a considerably large
region of the parameter space ($\xh1>0.2$ and $\tq>1.0$) with $4,000$
hr of observation, and we require approximately $8,000$ hr of
observation for a $5\sigma$ detection in this region of parameter
space.
 
The possibility of a 21-cm detection is quite favourable if the
quasar luminosity is of the order of 
$\dot{N}_{phs} = 3 \times 10^{57}\, {\rm sec^{-1}}$ or larger, and 
in much our analysis we use two different
 quasar luminosities $\dot{N}_{phs} = 3 \times 10^{57}\, {\rm
   sec^{-1}}$ and $8 \times 10^{57}\, {\rm sec^{-1}}$.  Here we
 present estimates of the SNR for bubble detection considering $1000$
 hr of GMRT observation around a known quasar. The right panels in
 Figure \ref{fig:contour_snr_rperp} show the peak SNR, which
 corresponds to the situation when the filter and the signal are
 exactly matched. And the left panels in Figure
 \ref{fig:contour_snr_rperp} show the corresponding quasar bubble size
 ($R_{\perp}$ in Mpc). An ionized bubble grows as $r(\tau)
   \propto (\tau \dot{N}_{phs}/\xh1)^{1/3}$
  in the early stages when $\tau \ll \tau_{rec}$ (eq. \ref{eq:growth})
  and saturates at,  
\begin{equation}
r(\tau)= r_s = 54 \,{\rm Mpc}\, \left( \frac{\dot{N}_{phs}}{3 \times 10^{57} {\rm sec}^{-1}}
\right)^{1/3} \,,
\end{equation}
for 
\begin{equation}
\tau > \tau_{rec} = 3.3 \times 10^7 \,{\rm yr}\, \xh1^{-1}\,.
\end{equation}
 We see that both these stages are distinctly
visible in the apparent size of the \HII bubble shown in Figure
\ref{fig:contour_snr_rperp}. Note that the value of $r_s$ where the growth saturates, is
independent of $\xh1$.

The SNR from an \HII bubble scales as $\xh1 R_b^{3/2}
\dot{N}_{phs}^{1/3}$ (Paper I), which implies that SNR $\propto \left(
\xh1 \tau \dot{N}_{phs} \right)^{1/2}$ when $\tau \ll \tau_{rec}$ and
SNR $\propto \xh1 \dot{N}_{phs}^{1/2}$ and independent of $\tau$ when
$\tau > \tau_{rec}$ .

We observe that the SNR contours are roughly
 hyperbolas in the $\xh1 - \tau_Q$ plane, and the hyperbolas flatten
 out at large $\tau$. We mainly focus on the
 SNR$=3$ contour, a $3\sigma$ (or higher) detection is possible in the
 parameter region to the right of this contour. We first consider the
 case where $(\dot{N}_{phs}/10^{57}\, {\rm sec^{-1}}) = 3$. We find
 that a $3\sigma$ detection is possible over a reasonably large
 region of the parameter space.  For a high neutral fraction ($0.6 \le
 \xh1$), it will be possible to detect an \HII bubble even if it is
 small ($R_{\perp} \approx 20 $ Mpc) and in an early stage of its
 growth $(\tau_Q/10^7 {\rm yr} \ge 0.5)$.  However, it will not be
 possible to detect the bubble if the quasar's age $(\tau_Q/10^7 {\rm
   yr})$ is less than $0.5$, even if the IGM is completely neutral
 outside the bubble.  In contrast, a $3 \sigma$ detection is possible
 for a low neutral fraction only if the bubble is very large and in a
 later stage of its evolution.  The lowest neutral fraction where a
 $3 \sigma$ detection is possible is $\xh1 \approx 0.2$. The quasar's
 age should be $\sim 3.0 \times 10^7$ yr (or more) for which the bubble
 radius is $R_{\perp} \sim 45 \, {\rm Mpc}$. For a quasar of this
 luminosity, a $3 \sigma$ detection is not possible if the neutral
 fraction is lower than $0.2$. We find that a $5\sigma$ detection also is
 possible for a reasonably large region of the parameter space roughly
 covering $\xh1>0.3$ and $\tq> 1$.

We next consider the case where $(\dot{N}_{phs}/10^{57}\, {\rm
  sec^{-1}}) = 8$.  The scaling behaviour discussed earlier leads us
to expect that $R_{\perp}$ and the SNR will increase by a factor of
$1.38$ and $1.63$ respectively if $\np$ is increased from $3$ to
$8$. Note however that, the scaling relations actually hold for the
bubble seen from the rest frame of the quasar, and we do no expect
them to be exactly valid for the bubble's apparent shape when the FLTT
is taken into account. The effect of FLTT is expected to be larger for
$\np=8$ where the bubble is predicted to be larger in comparison to
$\np=3$.

We see that there is a increase in the region of parameter space where
a $3\sigma$ detection is possible for the more luminous quasar.  The
smallest age for which a $3 \sigma$ detection is possible is now
reduced to $(\tau_Q/10^7 {\rm yr}) = 0.3$ compared to $0.5$ when
$(\dot{N}_{phs}/10^{57}\, {\rm sec^{-1}}) = 3$. This however requires
a completely neutral IGM. It is not possible to detect the bubble if
the quasar's age $(\tau_Q/10^7 {\rm yr})$ is less than $0.3$, even if
the IGM is completely neutral outside the bubble. At the other end, a
$3\sigma$ detection is possible if the neutral fraction is $\approx
0.1$, provided the quasar's age is $2.5 \times 10^7$ yr or more.  For
a quasar of this luminosity, it is not possible to detect bubbles if
the neutral fraction is less than $0.1$.

There are indications that the neutral fraction could have a value
$\xh1 \approx 0.5$ at $z = 8$ \citep{mitra}. For this neutral
  fraction a $3 \sigma$ detection is possible for $(\tau_Q/10^7 {\rm
    yr}) \ge 0.75$ and $0.5$ for $(\dot{N}_{phs}/10^{57}\, {\rm
    sec^{-1}}) = 3$ and $8$ respectively. The corresponding \HII
  bubble radii are $25$ and $30$ Mpc in these two cases
  respectively. We find that a $5\sigma$ detection is possible
  provided that the quasar's age exceeds $(\tau_Q/10^7 {\rm yr}) \ge
  2.0$ and $1.0$ respectively in these two cases.

We note that, the peak SNR does not increase very significantly when
we use the anisotropic filter instead of the spherical filter used in
earlier works. The anisotropic filter however has the advantage that
it parametrises the bubble in terms of $\xh1$ and $\tau_Q$ both of
which are physically relevant and interesting quantities in their own
right. We next consider the possibility of using matched filter
technique to observationally determine $\xh1$ and $\tau_Q$.
\begin{table}
\centering
\begin{tabular}{c|c|c}
\hline
\hline
 set & $\xh1$ & $\tau_Q/10^7$ \\
  &  & (yr) \\
\hline
 a0 & $0.75$ & $1.0$ \\
 b0 & $0.75$ & $1.8$ \\
 c0 & $0.75$ & $2.8$ \\
 d0 & $0.50$ & $1.2$ \\
 e0 & $0.50$ & $2.8$ \\
 f0 & $0.25$ & $2.8$ \\
\hline
\hline
\end{tabular}
\caption{ This tabulates the quasar and IGM  parameters  for which we
  have analytically considered parameter estimation.}
\label{tab:par}
\end{table}

\begin{figure*}
\psfrag{xh1}[c][c][1][0]{{\bf {\Large $\xh1$}}}
\psfrag{tq}[c][c][1][0]{{\bf {\Large $\tau_Q / 10^7\,$ yr}}}
\psfrag{a}[c][c][1][0]{{\bf {\LARGE  a0}}}
\psfrag{b}[c][c][1][0]{{\bf {\LARGE b0}}}
\psfrag{c}[c][c][1][0]{{\bf {\LARGE c0}}}
\psfrag{d}[c][c][1][0]{{\bf {\LARGE d0}}}
\psfrag{e}[c][c][1][0]{{\bf {\LARGE e0}}}
\psfrag{f}[c][c][1][0]{{\bf {\LARGE f0}}}
\includegraphics[width=.28\textwidth,
  angle=-90]{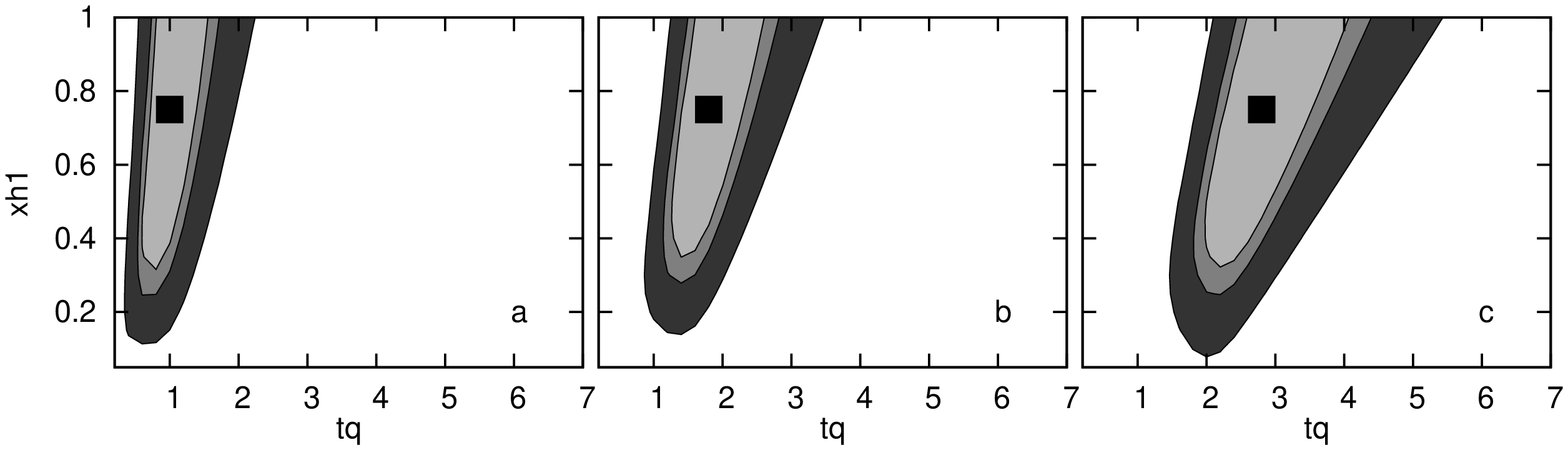}
\includegraphics[width=.28\textwidth,
  angle=-90]{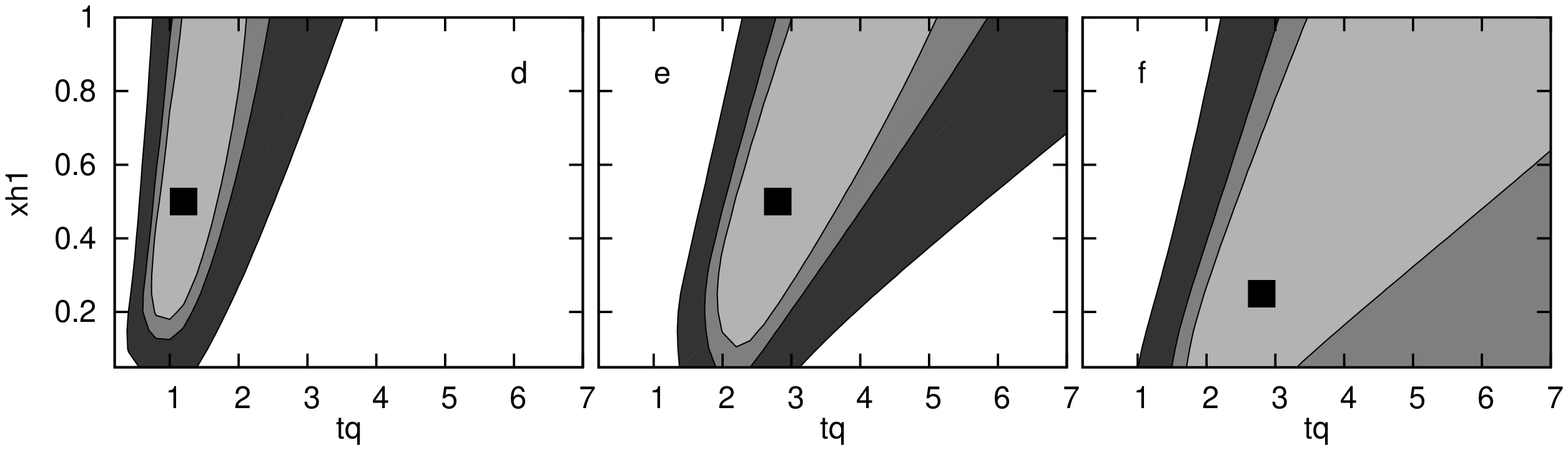}

\caption{This shows the uncertainty ($\Delta$ SNR $= 1$) in the
  estimated parameters ($\xh1$ and $\tau_Q$) for the parameter sets
  tabulated in Table \ref{tab:par}. The three shaded regions
  correspond to $1,000$, $4,000$ and $9,000$ hr of observations
  respectively. The quasar photon emission rate is $\np=3$ in all the
  panels.}
\label{fig:snr_contour1}
\end{figure*}
\begin{figure*}
\psfrag{xh1}[c][c][1][0]{{\bf {\Large $\xh1$}}}
\psfrag{tq}[c][c][1][0]{{\bf {\Large $\tau_Q / 10^7\,$ yr}}}
\psfrag{a}[c][c][1][0]{{\bf {\LARGE a0}}}
\psfrag{b}[c][c][1][0]{{\bf {\LARGE b0}}}
\psfrag{c}[c][c][1][0]{{\bf {\LARGE c0}}}
\psfrag{d}[c][c][1][0]{{\bf {\LARGE d0}}}
\psfrag{e}[c][c][1][0]{{\bf {\LARGE e0}}}
\psfrag{f}[c][c][1][0]{{\bf {\LARGE f0}}}
\includegraphics[width=.28\textwidth,
  angle=-90]{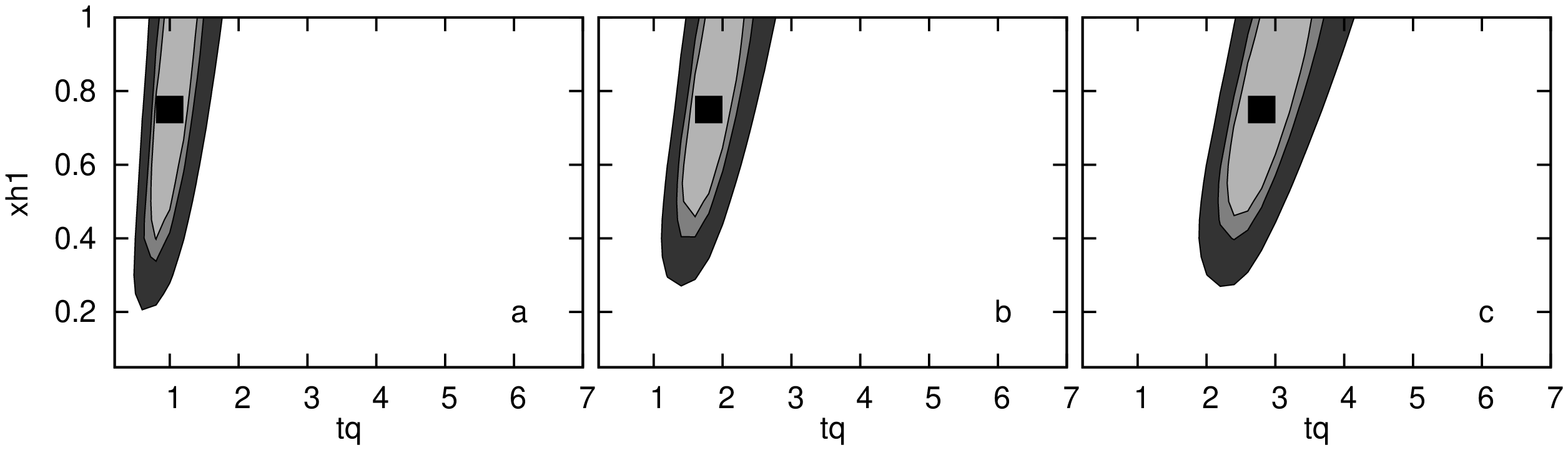}
\includegraphics[width=.28\textwidth,
  angle=-90]{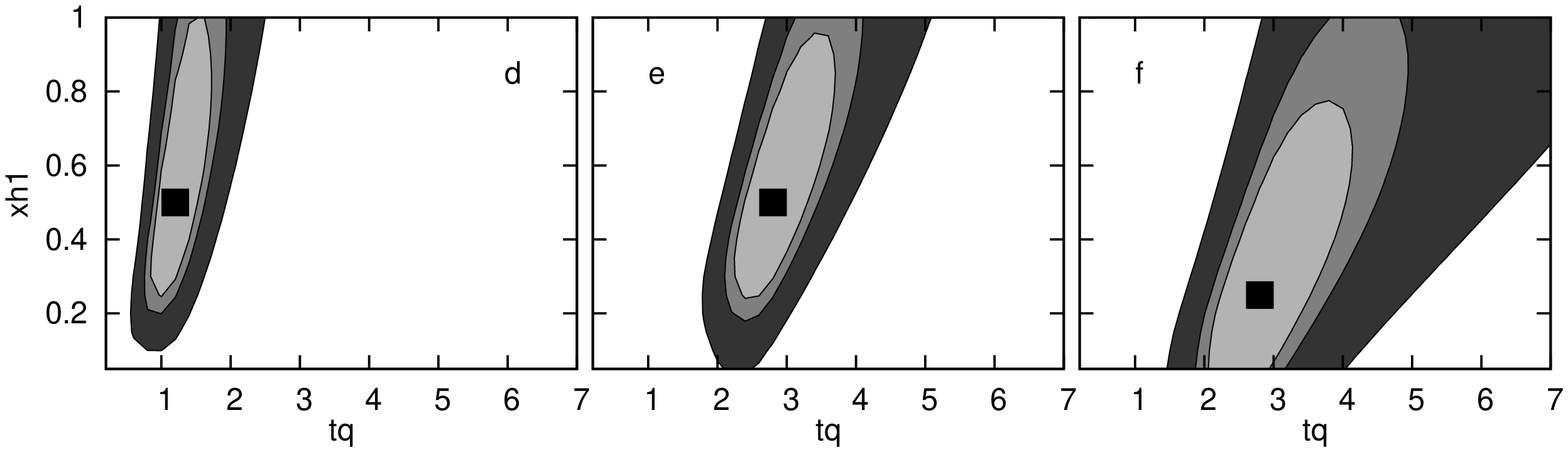}

\caption{Similar to Figure \ref{fig:snr_contour1} with 
  $\np=8$.}
\label{fig:snr_contour2}
\end{figure*}

\subsection{Parameter Estimation}
Once an \HII bubble has been detected in $1,000$ hr of GMRT
observation around a known quasar, it is natural to explore the
possibility of using this to estimate the quasar's age and the IGM
neutral fraction, $\tau_Q$ and $\xh1$ respectively. In most cases the
$1,000$ hr observation considered earlier is barely adequate for a
detection, and we do not expect very significant parameter estimation
from such an observation. Here we also consider deeper follow up
observations of $4,000$ and $9,000$ hr specifically carried out for the
purpose of parameter estimation once the \HII bubble has been
detected.

Parameter estimation is based on the idea that $\tau_Q$ and $\xh1$ are
uniquely imprinted in the \HII bubble's apparent size and shape
(Figure \ref{fig:shape_xh1}). Here we use $R_{\perp}$ (Figure
\ref{fig:contour_snr_rperp}) to quantify the bubble's apparent size,
and $\eta$ and $s$, introduced earlier, to quantify the bubble's
apparent shape.  Figure \ref{fig:contour_eta_s} shows the behaviour of
$\eta$ and $s$ for different values of the parameters $\tau_Q$ and
$\xh1$. Due to the FLTT the light reaching a present day observer was
emitted earlier from the back side of the bubble in comparison to the
light coming from the front side. For a bubble in the rapid stage of
growth ($\tau \ll \tau_{rec}$) the back side appears to have a much
smaller radius compared to its front side, causing the bubble to
appear elongated along the LoS. The centre of the apparent shape of
the bubble also shifts towards the observer. Both these features
are clearly seen in the parameter range $\tq \leq 1$ (Figure
\ref{fig:contour_eta_s}) where the bubble appears elongated ($\eta >
0$).  We see that both $\eta$ and $s$ are extremely sensitive to
$\tau_Q$ in this stage, however we do not see any significant $\xh1$
dependence. Though much of this region is outside the parameter range
where a $3\sigma$ detection is possible for $1000$ hr of GMRT
observation. For $\np=3$ and $8$, there are small regions with high
neutral fraction ($\xh1 \ge 0.65$) and $\tq > 0.5$ and $0.3$
respectively where it may be possible to detect an elongated bubble.

Most of the region ($\tq > 1$) of the $\xh1 - \tau_Q$ parameter space
where the \HII bubble is detectable, corresponds to the late stages of
its growth where the bubble appears compressed. The bubble is no
longer in the phase of rapid growth, and it appears mildly compressed
($\eta \leq 0$) along the LoS. There is however still a considerable
shift in the apparent centre of the bubble in this stage of growth.  In
this stage the anisotropy $\eta$ is sensitive to both $\xh1$ and
$\tau_Q$. The shift $s$ is mainly sensitive to $\tau_Q$ and largely
insensitive to $\xh1$ (Figure \ref{fig:contour_eta_s}). The bubble's
radius (Figure \ref{fig:contour_snr_rperp}) also is sensitive to both
$\xh1$ and $\tau_Q$ in this stage of the bubble's growth.  The period
$0.6 - 4.0 \times 10^7$ yr which is just after the rapid growth of the
bubble appears to be the most suitable for parameter estimation as the
apparent size and shape of the bubble are sensitive to both $\tau_Q$
and $\xh1$. The bubble's apparent shape and size are only weakly
sensitive to $\tau_Q$ beyond $\tq > 4.0$.  However, the bubble's size
remains sensitive to $\xh1$, and it may be possible to constrain this
parameter in the very late stage of growth.
\begin{figure*}
\psfrag{xh1}[c][c][1][0]{{\bf {\Large $\xh1$}}}
\psfrag{tq}[c][c][1][0]{{\bf {\Large $\tau_Q / 10^7\,$ yr}}}
\psfrag{C=20}[c][c][1][0]{{\bf {\Large $C=20$}}}
\psfrag{C=40}[c][c][1][0]{{\bf {\Large $C=40$}}}
\includegraphics[width=.3\textwidth,
  angle=-90]{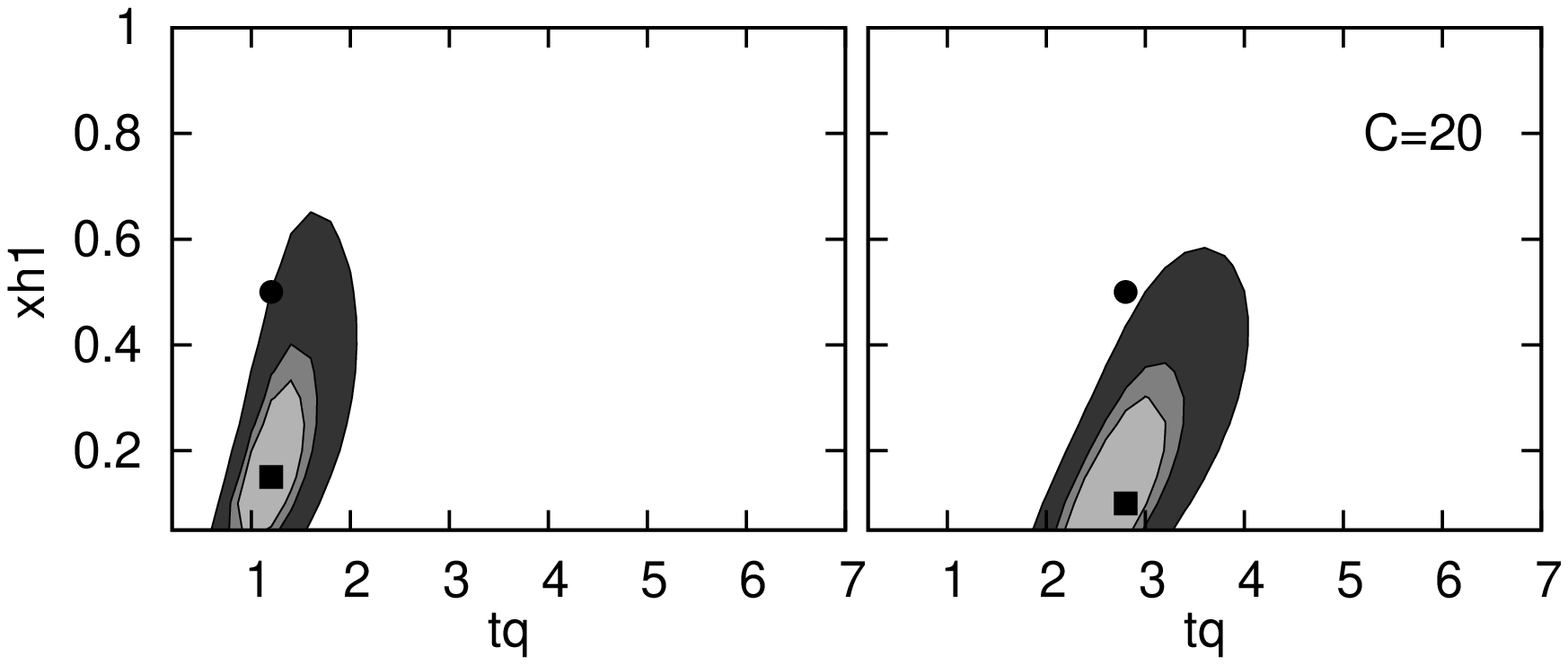}
\includegraphics[width=.3\textwidth,
  angle=-90]{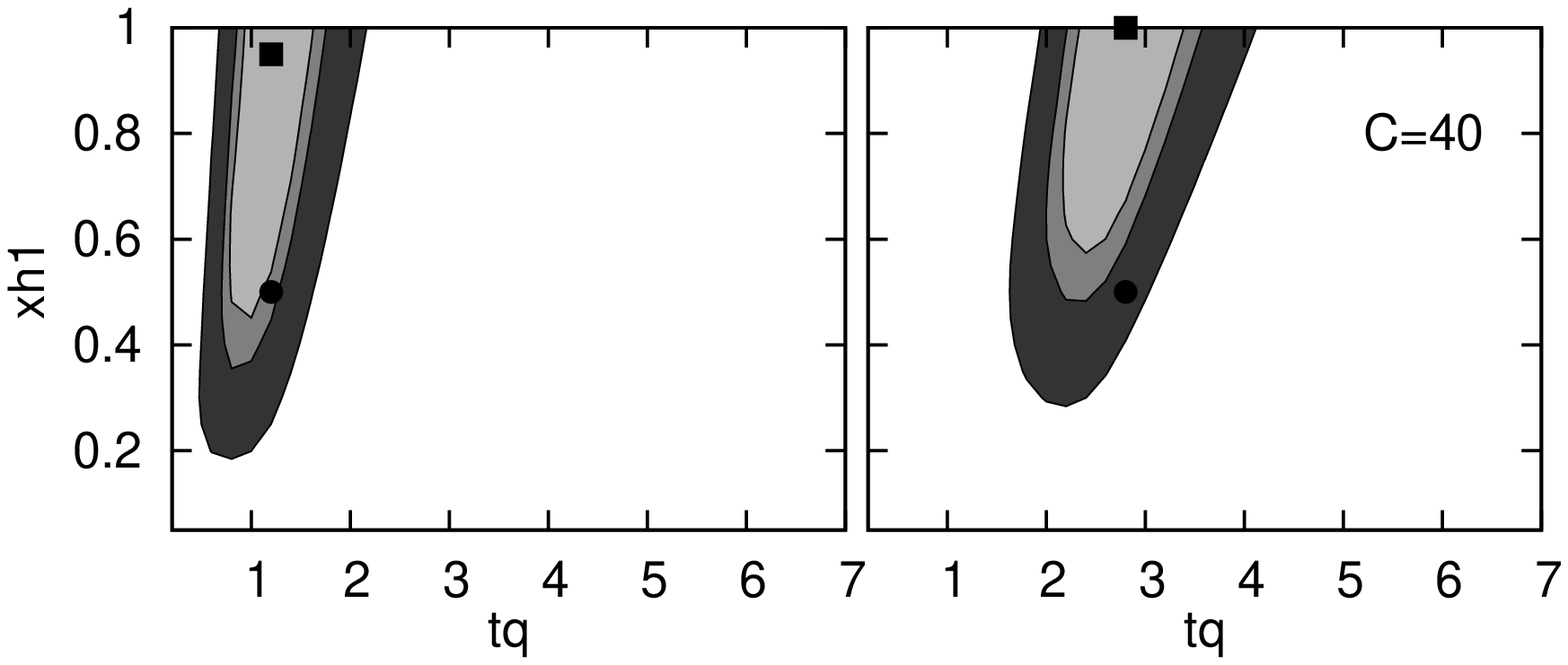}

\caption{This shows the estimated parameters in a situation where the
  value of $C$ assumed for the filter is different from the actual
  value in the IGM. Top and bottom panels correspond to the IGM having
  $C= 20$ and $40$ respectively. Left and right panels correspond to
  $\tq = 1.2$ and $2.8$ respectively. The filter assumes $C = 30$ in
  all cases. The bubble and the filter both have $\np=8$. The square
  in each panel shows the estimates from the search while the circle
  shows the actual input parameters. The shaded regions show the
  uncertainty ($\Delta$ SNR $= 1$) in the estimated parameters for
  $1,000$, $4000$ and $9,000$ hr of observation.}
\label{fig:diff_C}
\end{figure*}

Ideally we expect the SNR to peak when the parameters ($\xh1$,
$\tau_Q$) of the filter exactly match those of the bubble that is
actually present in the data. However the statistical fluctuations
will cause the position of the peak to shift introducing an
uncertainty in the parameter estimation. We use the criteria
$\Delta$SNR $=1$ to estimate the uncertainty in the estimated values
of the parameters $\xh1$ and $\tau_Q$. Table \ref{tab:par} shows the
sets of parameters for which we have considered parameter estimation
in this section.  The parameter sets a0, b0 and c0 have a high neutral
fraction $(\xh1=0.75)$, d0 and e0 have $\xh1=0.5$ and f0 has a low
neutral fraction $(\xh1=0.25)$. Figure \ref{fig:snr_contour1} and
\ref{fig:snr_contour2} show the uncertainties in parameter estimation
for all these cases assuming that $\np=3$ and $8$ respectively.

There are indications that the neutral fraction could have a value
$\xh1 \approx 0.5$ at $z = 8$ \citep{mitra}. We have considered two
sets of parameter values (d0 and e0) for this neutral
fraction. Considering $1,000$ hr of observation for the set d0, where
the bubble is in its early stage of growth ($\tq = 1.2$), we find that
it is possible to put both lower and upper limits on the quasar age
($0.5 \leq \tq \leq 3.5$) for a quasar with $\np = 3$. In this case no
limits on $\xh1$ can be placed with $1,000$ hr of observation. The
limits on $\tau_Q$ improves further and a lower limit can be placed on
$\xh1$ when the observation time is increased to $9,000$ hr. These
limits now are $1.0 \leq \tq \leq 2.0$ and $\xh1 \ge 0.2$.  We see
that there is a considerable improvement in parameter estimation if we
have a brighter quasar with $\np = 8$. We now have the limits $0.5
\leq \tq \leq 2.5$ and $\xh1 \ge 0.1$ for $1000$ hr of observation,
and $1.0 \leq \tq \leq 1.75$ and $\xh1 \ge 0.25$ for $9000$ hr of
observation.

We next consider the set e0 which corresponds to a later stage of growth
($\tq = 2.8$). We see that in this case the constraints on $\tau_Q$
and $\xh1$ are weaker compared to the set d where the quasar bubble is in an
earlier stage of its growth. Here it is possible to place only a lower
limit on $\tau_Q$ ($\tq > 1.5$) for a quasar with $\np = 3$ and
$1,000$ hr of observation. These limits become $2.0 \leq \tq \leq 5$
and $\xh1 \ge 0.15$ for $9,000$ hr of observation. In case we have a
brighter quasar ($\np = 8$), the limits are $1.75 \leq \tq \leq 5$ and
$2.25 \leq \tq \leq 3.75$ for $1,000$ and $9,000$ hr of observation
respectively, and $\xh1$ can only be constrained ( $0.25 \leq \xh1 $)
with $9,000$ hr of observation.
  
We next consider the possibility of a high neutral fraction
($\xh1=0.75$) for which a0, b0 and c0 respectively correspond to
progressively increasing quasar age. Comparing the results with those
for $\xh1=0.5$, we find that in it is possible to place tighter
constraints on $\tau_Q$ and $\xh1$ if the neutral fraction is
higher. We find that it is possible to place an upper and a lower
limit on $\tau_Q$, and a lower limit on $\xh1$ in all the cases that
we have considered.The constraints are tightest if the bubble is in
the early stage of its growth, and we have $0.5 \le \tq \le 2.25$ and $\xh1
\ge 0.1$ for $1,000$ hr of observation for a quasar with $\np=3$ and
$\tq=1.0$. The constraints are further improved if we have longer
observations or a brighter quasar.

We finally consider the set f0 which corresponds to a low neutral fraction
($\xh1=0.25$). In this case the bubble is not large enough for a
detection in the early stage of its growth (Figure
\ref{fig:contour_snr_rperp}), and we have considered $\tq = 2.8$ which
is in a later stage of its growth.  We find that the prospects of
constraining $\tau_Q$ and $\xh1$ are worse in comparison to the
situation where we have a high neutral fraction. For a quasar with
$\np=3$ it is not possible to constrain $\xh1$ even with $9,000$ hr of
observation, and it is possible to place only a lower limits $\tq \ge
1$ and $ \tq \ge 1.75$ with $1,000$ and $9,000$ hr respectively.  For
a bright quasar $(\np=8)$, it is possible place constraints $2 \le
\tq \le 4$ and $\xh1 \le 0.75$ with $9,000$ hr of observation.

In summary, we find that  the  situation is most favourable for 
constraining $\tau_Q$ and $\xh1$ if we can detect a very luminous
quasar in the early phase of its growth and also in the early stage of
reionization when the neutral fraction is high.
  
As mentioned earlier in this section the anisotropy in the \HII
bubble's shape is determined by three parameters $\dot{N}_{phs}/C$,
$\xh1/C$ and $\tau_Q$. We have assumed $C = 30$ for both the filter
and the bubble. It is quite possible that the actual clumping factor
is different from $C = 30$. This will introduce a systematic
uncertainty in parameter estimation. We consider two situations where
the actual value of $C$ for the bubble is different ($C = 20$ and
$40$) than the filter ($C = 30$) but the bubble and the filter both
have same photon emission rate ($\np = 8$) in their models. Top and
bottom panels of Figure \ref{fig:diff_C} (corresponding to $C = 20$
and $40$ respectively) show the estimated values of parameters in
these two different situations. We find that errors in $C$ mainly
affect the $\xh1$, $\tau_Q$ is largely unaffected by this. In a
situation where the $C$ in the IGM of the bubble is less ($C = 20$)
than what is assumed for the filter ($C = 30$), the filter severely
underestimates $\xh1$ and this underestimation becomes more in the
late stage ($\tq = 2.8$) of the growth than the early stage ($\tq =
1.2$). On the other hand when the $C$ in the bubble's IGM is more ($C
= 40$) than that of the filter, the $\xh1$ is severely overestimated
and this overestimation is more in the late stage ($\tq = 2.8$) of the
bubble's growth.

Similarly $\dot{N}_{phs}$ estimated from the spectrum of the quasar
could be off from the actual value. It will also have similar effects
as the errors in $C$ in parameter estimations.

\section{Simulating the ionization map}
\begin{figure*}
\psfrag{Mpc}[c][c][1][0]{{\bf {\Large Mpc}}}
\psfrag{p=180}[c][c][1][0]{\textcolor{white}{{\bf {\large {\boldmath
          $\phi=180^{\circ}$}}}}}
\psfrag{p=135}[c][c][1][0]{\textcolor{white}{{\bf {\large {\boldmath
          $\phi=135^{\circ}$}}}}}
\psfrag{p=90}[c][c][1][0]{\textcolor{white}{{\bf {\large {\boldmath
          $\phi=90^{\circ}$}}}}}
\psfrag{p=60}[c][c][1][0]{\textcolor{white}{{\bf {\large {\boldmath
          $\phi=60^{\circ}$}}}}}
\psfrag{p=45}[c][c][1][0]{\textcolor{white}{{\bf {\large {\boldmath
          $\phi=45^{\circ}$}}}}}
\psfrag{p=0}[c][c][1][0]{\textcolor{white}{{\bf {\large {\boldmath
          $\phi=0^{\circ}$}}}}}
\psfrag{t=0.30}[c][c][1][0]{\textcolor{white}{{\bf {\large {\boldmath
          $0.30 \times 10^7\,{\rm yr}$}}}}}
\psfrag{t=0.46}[c][c][1][0]{\textcolor{white}{{\bf {\large {\boldmath
          $0.46 \times 10^7\,{\rm yr}$}}}}}
\psfrag{t=0.91}[c][c][1][0]{\textcolor{white}{{\bf {\large {\boldmath
          $0.91 \times 10^7\,{\rm yr}$}}}}}
\psfrag{t=1.40}[c][c][1][0]{\textcolor{white}{{\bf {\large {\boldmath
          $1.40 \times 10^7\,{\rm yr}$}}}}}
\psfrag{t=1.64}[c][c][1][0]{\textcolor{white}{{\bf {\large {\boldmath
          $1.64 \times 10^7\,{\rm yr}$}}}}}
\psfrag{t=2.00}[c][c][1][0]{\textcolor{white}{{\bf {\large {\boldmath
          $2.00 \times 10^7\,{\rm yr}$}}}}}
\includegraphics[width=1.0\textwidth, angle=0]{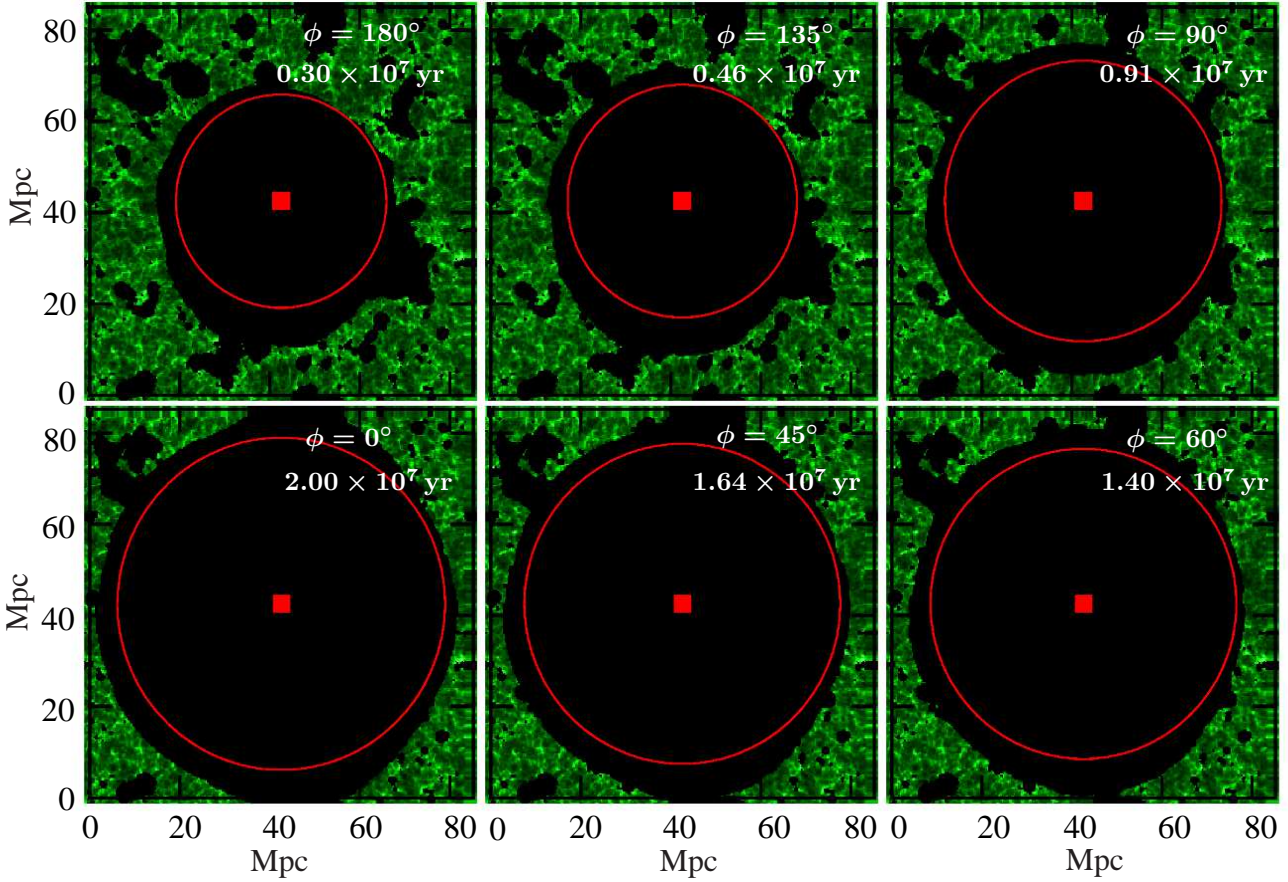}
\caption{This shows a simulated quasar \HII bubble with $\np=1.3$ and
  $\xh1 = 0.5$ at different stages of its growth, as seen from the
  quasar's rest frame. The square in each panel shows the location of
  the quasar. The expected spherical size is shown by the
  circle. Panels are labelled with the corresponding quasar age and
  $\phi$ is the angle with the LoS as defined in Figure \ref{fig:shape}
  and eq. (\ref{eq:shape}).}
\label{fig:growth}
\end{figure*}

For our analytic estimates in the previous section we have considered
that the \HI density is uniform in the IGM around the quasar and there
is no other ionizing source in the IGM except the target quasar. It is
expected that in reality the IGM \HI density will not be uniform and
the stellar sources in the vicinity of the quasar will also contribute
in ionizing the IGM. To study the detectability and parameter
estimation in a more realistic situation when all these effects are
taken into account we have simulated the ionization map around the
quasar using a semi-numerical formalism.
 
We have implemented the semi-numerical formalism proposed by
\citet{choudhury09b} for generating the ionization field at a given
redshift. This uses an excursion-set formalism as introduced by
\citet{furlanetto1}.

It is currently believed that stars residing in galaxies are the major
source of photons to reionize the universe
\citep{choudhury06,choudhury09a}. The early stages of reionization are
driven by stars within halos of mass $\sim 10^8 \,h^{-1}\,
M_{\odot}$. As the IGM becomes ionized, star formation within the
smaller halos ($M < 10^9 \,h^{-1}\, M_{\odot}$) is inhibited because
of radiative feedback. Hence, it is sufficient to only include halos
of mass $M \geq 10^9 \,h^{-1}\, M_{\odot}$ to simulate  
the final stages of reionization. The smallest halo that is resolved
in the simulation should be of mass $M \leq 10^9 \,h^{-1}\,
M_{\odot}$. If at least $10$ particles are required to constitute the
smallest halo, then the particle mass should be $\leq 10^8 \,h^{-1}\,
M_{\odot}$. This decides the mass resolution of our simulation.

We have generated the dark matter distribution at $z = 8$ using a
Particle Mesh Nbody code developed by \citet{Bharadwaj04}.  The volume
of the simulation is constrained by the $16$ Gigabytes of memory
available in our computer. We perform our simulation in a periodic box
of size $85.12$ Mpc (comoving) with $1216^3$ grid points and   $608^3$
particles, with a mass resolution $M_{part} = 7.275 \times
10^7\,h^{-1}\,M_{\odot}$.

We identify  halos within the simulation box using a standard
Friend-of-Friend algorithm \citep{davis}, with a fixed linking length
$b=0.2 $ (in units of mean inter particle distance) and minimum halo
mass $ = 10 M_{part}$.  We also compare the comoving number density of
halos per unit logarithmic mass interval $dn/d\,ln\,M$ with the
theoretical mass function at $z=8$ as predicted by \citet{sheth} using 
the fitting 
function adopted from \citet{jenkins}. A good agreement is found over
a wide mass range $10^9 \lesssim M/(h^{-1} M_{\odot}) \lesssim 5
\times 10^{11}$. The lower mass limit corresponds to the smallest halo
mass.

The relation between the ionizing luminosity of a galaxy and its 
properties is not well known from observations. In the semi-numerical
formalism adopted here, we assume that the ionizing luminosity from
a galaxy is proportional to the mass of its halo. The number of
ionizing photons contributed by a halo of mass $M$ is given by,
\begin{equation}
N_{\gamma}(M) = N_{ion} \frac{M}{m_H}\,,
\label{eq:nion} 
\end{equation}
where $m_H$ is the mass of a hydrogen atom and $N_{ion}$ is a
dimensionless constant. The value of $N_{ion}$ is tuned so as to
achieve the value $\xh1$ desired in the simulation. In the
semi-numerical formalism, a region is said to be ionized if the
average number  of photons reaching there exceeds average neutral 
hydrogen density at that point. We have used a $256^3$ grid with a grid
spacing of $0.3325$ Mpc for simulating the ionization maps.

The semi-numerical formalism provides a snapshot of the ionized IGM at
a fixed instant of time $t_s$. We now briefly discuss how we have
incorporated a quasar as an extra ionizing source in the
simulation. The quasar is assumed to have ionizing luminosity
$\dot{N}_{phs}$ and age $\tau$ at the instant $t_s$. We have
identified the most massive halo ($M \sim 5 \times 10^{11} h^{-1}
M_{\odot}$) in the simulation as the quasar's location. We also shift
the entire box with periodic boundary condition so as to bring the
quasar into the centre of the field of view (FoV). In order to
incorporate the quasar's \HII bubble it is necessary to provide the
simulation with a $N_{\gamma,QSO}$, the number  of photons corresponding
to the quasar at the instant $t_s$. We have used,
\begin{equation}
N_{\gamma,QSO} = \frac{4}{3} \pi \langle n_{H} \rangle \xh1
r^3(\tau) \,,
\label{eq:qso}
\end{equation}
where $r(\tau)$ is determined using eq. (\ref{eq:growth}), {\it
  i.e.} the \HII bubble would have a comoving radius $r(\tau)$ if
the hydrogen is uniformly distributed and there were no ionizing
sources other than the quasar.

The quasar bubble actually produced in the simulation differs from
this because of inhomogeneities in the hydrogen distribution and the
presence of other ionizing sources. Further the density dependent
non-uniform recombinations also change the bubble's size and shape in
the cases where this effect is included. In general, we find the
simulated \HII bubbles are bigger than predicted by the
eq. (\ref{eq:growth}), mainly because of the contributions of other
ionizing sources near the quasar. Figure \ref{fig:growth} provides a
visual impression of a suite of simulations with $\np = 1.3$, $\xh1 =
0.5$ and different values of $\tau$. We notice that there is a
particularly large deviation from the expected bubble size at the
early stages of the bubble's growth. The actual bubble is much larger
than that expected from the $N_{\gamma,QSO}$ that we have assigned to
the quasar. The massive halos, which are the only other ionizing
sources besides the quasar, are preferentially clustered in the
vicinity of the quasar which is located in the most massive halo in
the simulation. The ionizing photons from these halos outnumber those
from the quasar in the early stages, and consequently the actual
bubble is much larger than expected. This discrepancy persists even in
the later stages, however it is not so pronounced. A similar
observation has been reported by \citet{datta12} in their radiative
transfer simulations.

\begin{figure}
\psfrag{Mpc}[c][c][1][0]{{\bf {\Large Mpc}}}
\includegraphics[width=.5\textwidth, angle=-0]{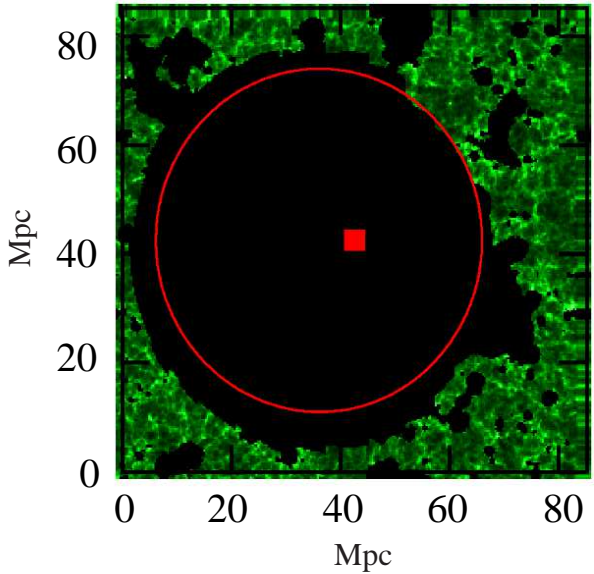}
\caption{This shows the apparent shape of the \HII bubble around a
  quasar after the snapshots at different stages of its growth (Figure
  \ref{fig:growth}) have been stacked together. The square shows the
  location of the quasar. The curve shows the expected apparent shape
  of the bubble. The observer is on the left of the figure.}
\label{fig:ani_shape}
\end{figure}

A distant observer sees different parts of the \HII bubble in
different stages of its growth. We have assumed that the dark matter
distribution and the global $\xh1$ do not change within the look back
time across the bubble. We have used eq. (\ref{eq:growth}) and
(\ref{eq:shape}) to determine $r$ as a function of $\phi$. We then
choose sections each through a different  member of the suite of
simulations (Figure \ref{fig:growth}). Each section corresponds to a
different value of $r$ and $\tau$ as determined by
eq. (\ref{eq:growth}) and (\ref{eq:shape}). These sections are stacked
together to produce the bubble's apparent shape as seen by the distant
observer. Figure \ref{fig:ani_shape} shows the apparent shape
generated by stacking   sections selected from the Figure
\ref{fig:growth}. 

Our simulations spans a redshift interval $\Delta z = 0.28$ along the
LoS. We have made a simplifying assumption that the dark matter
distribution and the variation in $\xh1$ across our simulation box can
be ignored. Analytic estimates (for details see Appendix A and Results
of Paper II) show that the effects ignored here make a $5\%$ or less
contribution. Further, simulations \citep{datta11} also reveal
similar findings, justifying the assumptions made here.

The GMRT FoV at $151$ MHz has a full width at half maxima (FWHM) $=
2.28^{\circ}$ while our simulation box only subtends an angle
$0.53^{\circ}$ on the sky and $4.8$ MHz along the LoS at redshift $z =
8$. Thus the simulation box will not be able to replicate the full
GMRT FoV and can accommodate a bubble of maximum radius $\sim 40$ Mpc,
which subtends $\sim 0.5^{\circ}$. \citet{maselli07} predict the
comoving radius of quasar generated \HII regions to be $\sim 45$ Mpc
at $z = 6.1$ with $\xh1 =0.1$. The size is expected to be less at $z =
8$. Our simulation box is thus large enough to host \HII bubbles in
the relevant size range at this redshift. The simulation adequately
replicate the \HI fluctuations in the vicinity of the bubble, however
the \HI fluctuations at large angular separation from the bubble's
centre are missing. The \HI fluctuations are some what underestimated
as a consequence (see \citealt{datta3} for more details).

The simulated \HI maps were converted to GMRT visibilities which were
then used for matched filter bubble detection following the
procedure detailed in \citet{datta3}.

\section{Results}
\begin{table}
\centering
\begin{tabular}{c|c|c|c}
\hline
\hline
 set & $\dot{N}_{phs}/10^{57}$ & $\xh1$ & $\tau_Q/10^7$ \\
  & (${\rm sec^{-1}}$) & & (yr) \\
\hline
 A & $1.3$ & $0.50$ & $2.00$ \\
 B & $1.3$ & $0.50$ & $3.00$ \\
 C & $1.3$ & $0.75$ & $1.50$ \\
 D & $1.3$ & $0.75$ & $3.50$ \\
 E & $3.0$ & $0.50$ & $0.75$ \\
 F & $3.0$ & $0.75$ & $0.50$ \\
\hline
\hline
\end{tabular}
\caption{ This tabulates the quasar and IGM  parameters  for
which the \HII bubbles have been simulated.}
\label{tab:sim}
\end{table}

\begin{figure*}
\psfrag{Mpc}[c][c][1][0]{{\bf{\LARGE Mpc}}}
\psfrag{A}[c][c][1][0]{{\bf{\LARGE {\textcolor{white} A}}}}
\psfrag{B}[c][c][1][0]{{\bf{\LARGE {\textcolor{white} B}}}}
\psfrag{C}[c][c][1][0]{{\bf{\LARGE {\textcolor{white} C}}}}
\psfrag{D}[c][c][1][0]{{\bf{\LARGE {\textcolor{white} D}}}}
\psfrag{E}[c][c][1][0]{{\bf{\LARGE {\textcolor{white} E}}}}
\psfrag{F}[c][c][1][0]{{\bf{\LARGE {\textcolor{white} F}}}}
\includegraphics[width=1.0\textwidth,angle=-0]{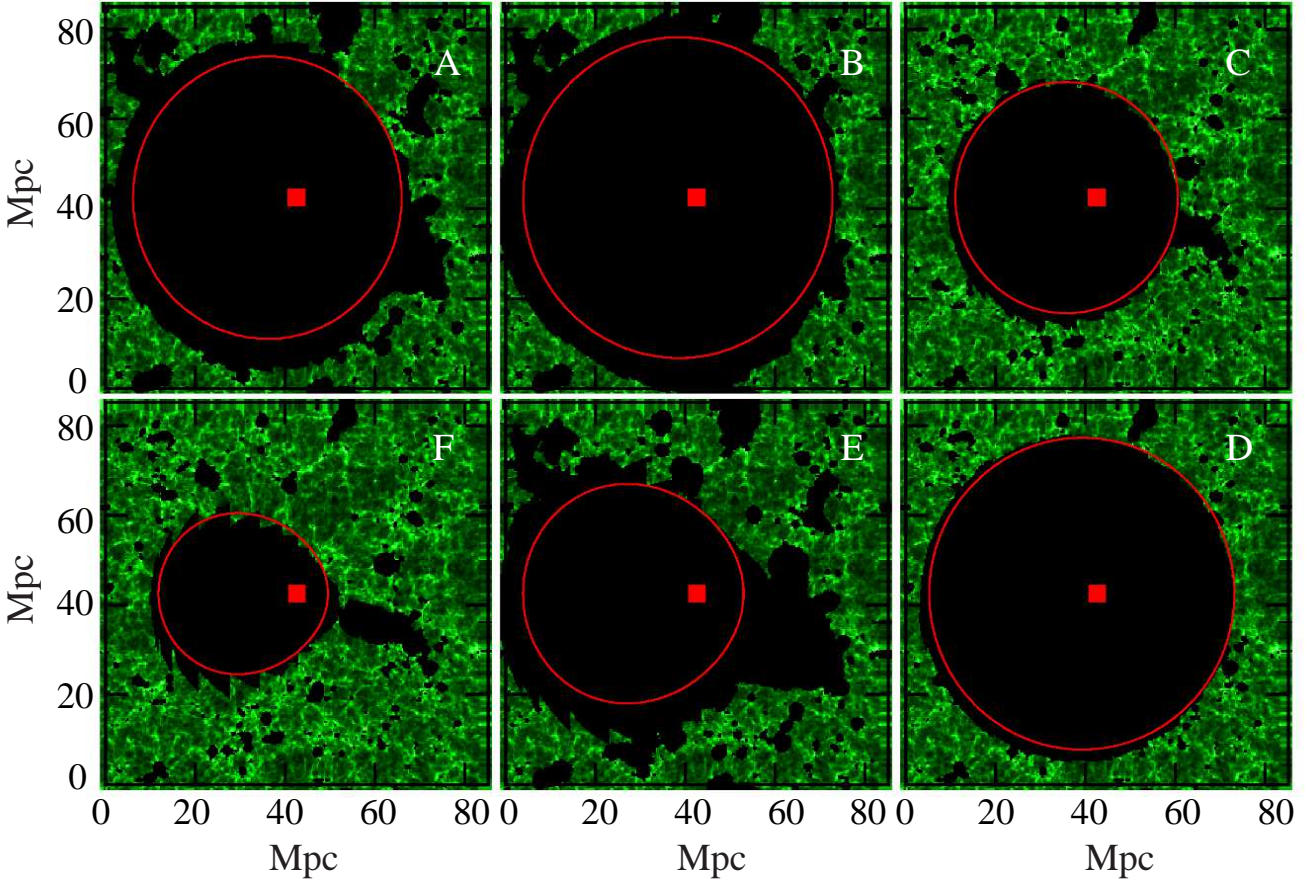}
\caption{This  shows the simulated bubbles for different sets of
  quasar and IGM parameters as tabulated in Table \ref{tab:sim}
  considering uniform recombination in the IGM. The square in each
  panel represents the location of the quasar. The curve shows the 
  expected bubble   shape. In all panels the  observer is to  the
  left side of the box.}
\label{fig:diff_eta_u}
\end{figure*}
\begin{figure*}
\psfrag{Mpc}[c][c][1][0]{{\bf{\LARGE Mpc}}}
\psfrag{A}[c][c][1][0]{{\bf{\LARGE {\textcolor{white} A}}}}
\psfrag{B}[c][c][1][0]{{\bf{\LARGE {\textcolor{white} B}}}}
\psfrag{C}[c][c][1][0]{{\bf{\LARGE {\textcolor{white} C}}}}
\psfrag{D}[c][c][1][0]{{\bf{\LARGE {\textcolor{white} D}}}}
\psfrag{E}[c][c][1][0]{{\bf{\LARGE {\textcolor{white} E}}}}
\psfrag{F}[c][c][1][0]{{\bf{\LARGE {\textcolor{white} F}}}}
\includegraphics[width=1.\textwidth,angle=-0]{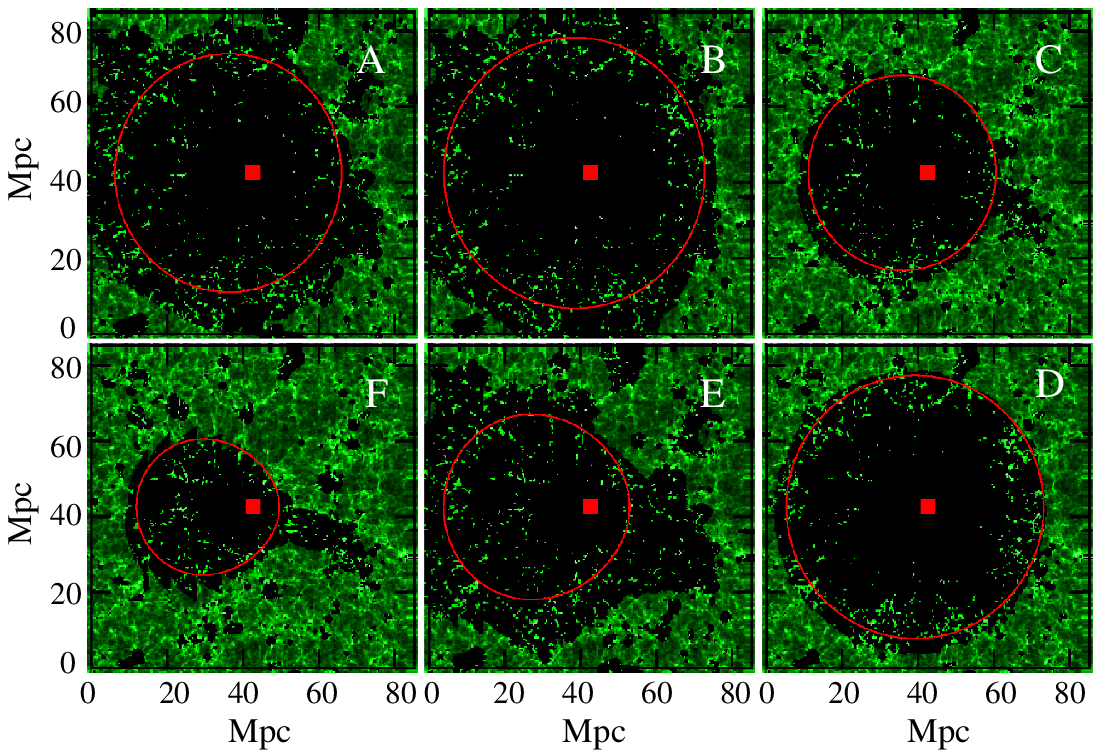}
\caption{Similar to Figure \ref{fig:diff_eta_u} considering
  non-uniform recombination in the IGM.}
\label{fig:diff_eta_n}
\end{figure*}

We have created six different realizations of the dark matter
distribution using the PM Nbody code. These were used to generate six
independent realizations of the quasar bubble for each of the
parameter sets given in Table \ref{tab:sim}. We have seen that it is
possible to constrain different regions of the $\xh1 - \tau_Q$
parameter space by observing quasars in different stages of
growth. Ideally we would like to simulate bubbles in different stages
of the growth for a variety of $\dot{N}_{phs}$ and $\xh1$.  However
our choice of parameters are restricted by the condition that the
bubble should be small enough ($< 40$ Mpc) to be contained within the
simulation box, while it should also be large enough for a $3\sigma$
detection (Figure \ref{fig:contour_snr_rperp}).  Figure
\ref{fig:diff_eta_u} shows the simulated \HII bubble for each set of
the parameters listed in Table \ref{tab:sim}.  The images shown in the
figure all correspond to the same realization, with uniform
recombination. Three sets of simulations A, B and E have
$\xh1=0.5$. The sets C, D and F have a higher neutral fraction
($\xh1=0.75$). We are unable to simulate any situation with a low
$\xh1$ because the bubble size would have to be larger than our box
for a detection to be possible. The sets A, B, C and D have a quasar
luminosity $\np=1.3$ which is the same as the quasar detected by
\citet{mortlock11}. This luminosity is relatively low in comparison to
the values that we have considered earlier. The quasar has to be
observed at a late stage for a detection to be possible, and we expect
a large bubble. The bubble centre will be shifted from the quasar for
A and C where the quasar is somewhat younger with $\tq=2.0\, {\rm
  and}\, 1.5$ respectively compared to B and D which have $\tq = 3.0
\, {\rm and}\, 3.5$ respectively. The bubbles are also expected to be
compressed in sets A and C, which is distinctly visible in these
images. The bubble is expected to be nearly spherical in sets B and D.

We see that the simulated bubbles are in reasonable agreement with
what is expected, though the simulated bubbles are nearly always
larger. This discrepancy arrises due to the extra ionizing photons
contributed by the halos located within the bubble. The number of
ionizing photons contributed by these halos is proportional to the
ionization fraction ($1 - \xh1$). Consequently the discrepancy between
the expected bubble size and that obtained in the simulations is
larger for $\xh1 =0.5$ in comparison to $\xh1 = 0.75$. Further, the
bubble's boundary is distorted because the bubble gets merged with the
ionized regions produced by halos located in the periphery of the
bubble.  These distortions are particularly important when the halos
located near the bubble's boundary contribute a significant fraction
of the total ionizing photons within the bubble. We thus expect these
distortions to be relatively important when the bubble is in the early
stage of growth and we have a low neutral fraction, whereas we expect
the distortions to be relatively less important when the bubble is in
a later stage and in an IGM with a higher $\xh1$.

The sets E and F have $\np=3$, and  the quasar's ages are smaller
($\tq = 0.75\, {\rm and}\, 0.50$) in comparison to
 sets A,B,C and D discussed earlier.  We expect the bubbles to
be elongated along the LoS in these two sets, and  this is
clearly visible in Figure \ref{fig:diff_eta_u}. However, the bubble's
shape is severely distorted in E where 
the neutral fraction is less compared to F. We expect such distortions
to severely affect parameter estimation.  We see that the distortion
present in E also persists for F, however the relative contribution is
smaller. We may expect the impact of this distortion on parameter
estimation to be less severe in F as compared to E. 

Figure \ref{fig:diff_eta_n} shows images from simulations with
inhomogeneous recombination. We see that the bubbles are larger in
comparison to the situation with uniform recombination.  It is
possible to understand this by noting that in these models the dense
regions in the IGM remain neutral through enhanced recombination, and
only the low density regions are ionized. Consequently, a larger
volume has to be ionized in order to achieve a particular value of the
mass averaged neutral fraction $\xh1$. Further, the contrast between
the bubble and the IGM outside is reduced, and the bubble's boundary
is more distorted.

\begin{figure*}
\psfrag{Mpc}[c][c][1][0]{{\bf{\LARGE Mpc}}}
\includegraphics[width=0.8\textwidth,angle=-0]{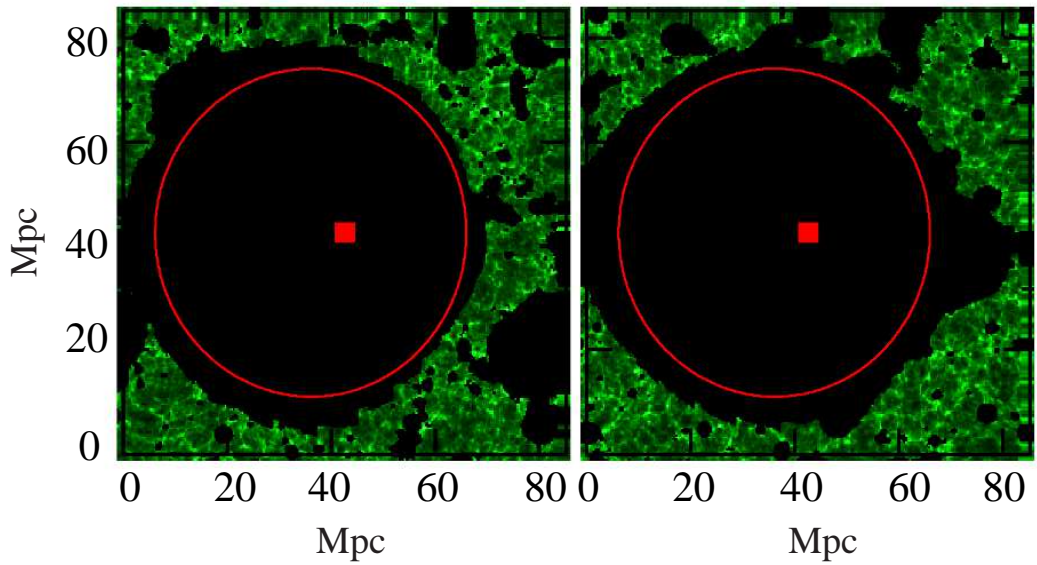}
\caption{This  shows two different realizations of the simulated
  \HII bubbles having the same set of parameters (set A). The effect
  of the   photon contribution from neighbouring halos is severe in
  the right  panel as  compared to the bubble in the left panel. The
  curve shows the   expected bubble   shape. In both panels the
  observer is to the left  of the box.}
\label{fig:diff_rel}
\end{figure*}

For a fixed set of parameters, we expect the simulated bubble to
differ from one realization to the next.  Figure \ref{fig:diff_rel}
shows two different realizations of the simulated bubble for the
parameter set A in addition to the realization shown in Figure
\ref{fig:diff_eta_u}.  We see that in all the realizations the bubble
appears larger than expected. The distortions, we notice, can vary
significantly from one realization to the next.  In particular, the
right panel of Figure \ref{fig:diff_rel} shows a realization where an
extraneous ionization source is so aligned that it causes the ionized
bubble to appear elongated along the LoS. Recollect that the bubble is
expected to appear compressed in set A. Such distortions are a source
of concern for parameter estimation.
\begin{figure*}
\psfrag{Simulation}[c][c][1][0]{{\bf{\Large Simulation}}} 
\psfrag{SNR}[c][c][1][0]{{\bf{\Large SNR}}} 
\psfrag{A}[c][c][1][0]{{\bf{\Large A}}}
\psfrag{B}[c][c][1][0]{{\bf{\Large B}}}
\psfrag{C}[c][c][1][0]{{\bf{\Large C}}}
\psfrag{D}[c][c][1][0]{{\bf{\Large D}}}
\psfrag{E}[c][c][1][0]{{\bf{\Large E}}}
\psfrag{F}[c][c][1][0]{{\bf{\Large F}}}
\includegraphics[width=.4\textwidth, angle=-90]{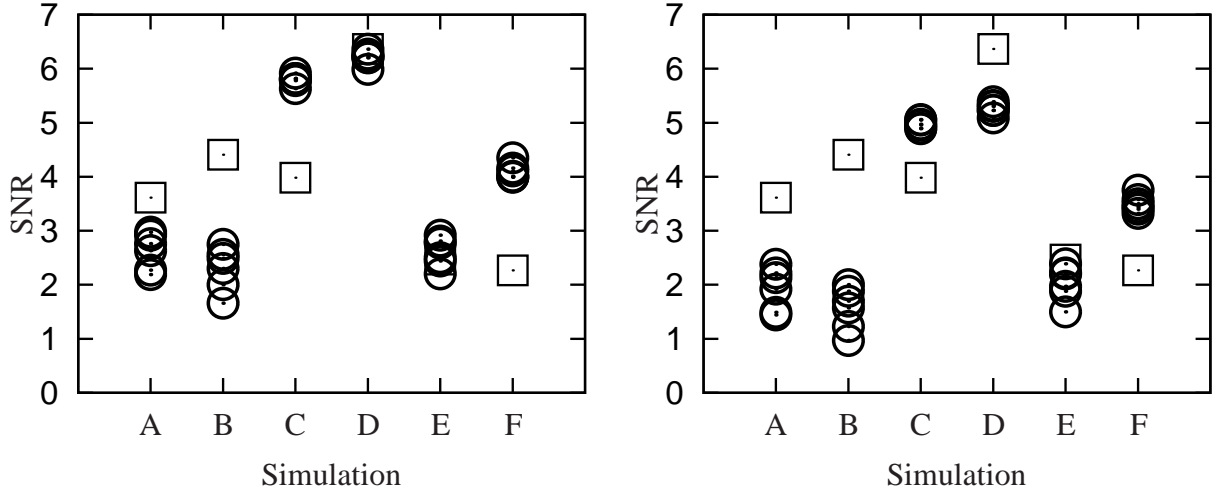}
\caption{This shows the SNR for $1,000$ hr of observation for the
  different sets of parameters in Table \ref{tab:sim}.  The squares
  indicates the analytic estimates and the circles represent the SNR
  estimated from different realizations of the simulations. Left and
  right panels correspond to simulations with uniform and non-uniform
  recombinations respectively.}
\label{fig:SNR}
\end{figure*}
\begin{figure*}
\psfrag{Simulation}[c][c][1][0]{{\bf{\Large Simulation}}} 
\psfrag{Rperp}[c][c][1][0]{{\bf{\Large $R_{\perp}$ in Mpc}}} 
\psfrag{A}[c][c][1][0]{{\bf{\Large A}}}
\psfrag{B}[c][c][1][0]{{\bf{\Large B}}}
\psfrag{C}[c][c][1][0]{{\bf{\Large C}}}
\psfrag{D}[c][c][1][0]{{\bf{\Large D}}}
\psfrag{E}[c][c][1][0]{{\bf{\Large E}}}
\psfrag{F}[c][c][1][0]{{\bf{\Large F}}}
\includegraphics[width=.4\textwidth, angle=-90]{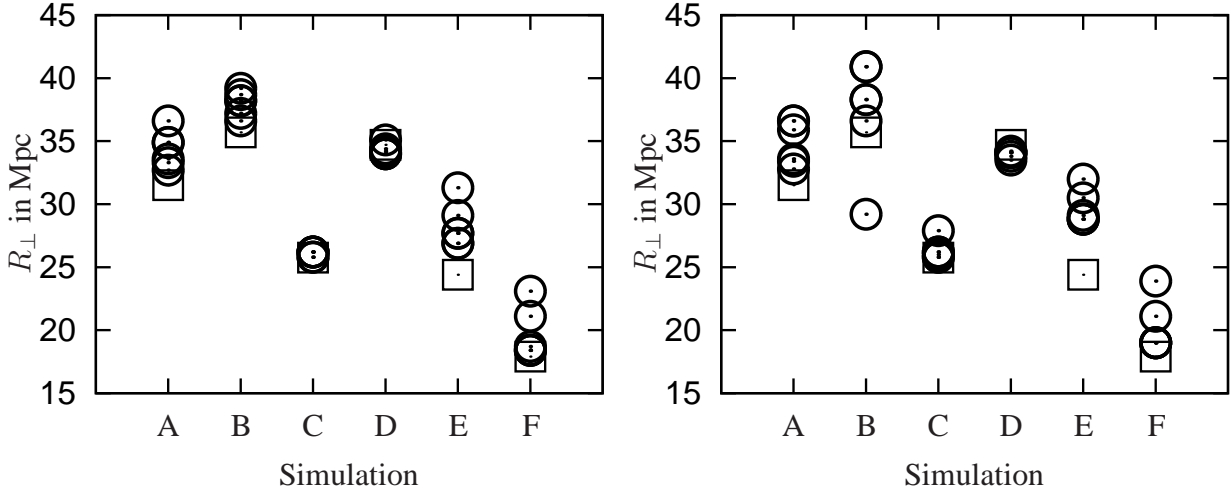}
\caption{This shows the bubble size estimated from matched filter
  search on simulated \HII bubbles for the different sets of
  parameters in Table \ref{tab:sim}. The squares indicate the
  analytic estimates of $R_{\perp}$ and the circles represent the same
  estimated from different realizations of the simulations. Left and
  right panels correspond to the simulations with uniform and non-uniform
  recombinations respectively.}
\label{fig:rperp}
\end{figure*} 
\begin{figure*}
\psfrag{xh1}[c][c][1][0]{{\bf{\Large $\xh1$}}} 
\psfrag{tq}[c][c][1][0]{{\bf{\Large $\tau_Q / 10^7\,$ yr}}} 
\psfrag{A}[c][c][1][0]{{\bf{\LARGE A}}}
\psfrag{B}[c][c][1][0]{{\bf{\LARGE B}}}
\psfrag{C}[c][c][1][0]{{\bf{\LARGE C}}}
\psfrag{D}[c][c][1][0]{{\bf{\LARGE D}}}
\psfrag{E}[c][c][1][0]{{\bf{\LARGE E}}}
\psfrag{F}[c][c][1][0]{{\bf{\LARGE F}}}
\includegraphics[width=.3\textwidth, angle=-90]{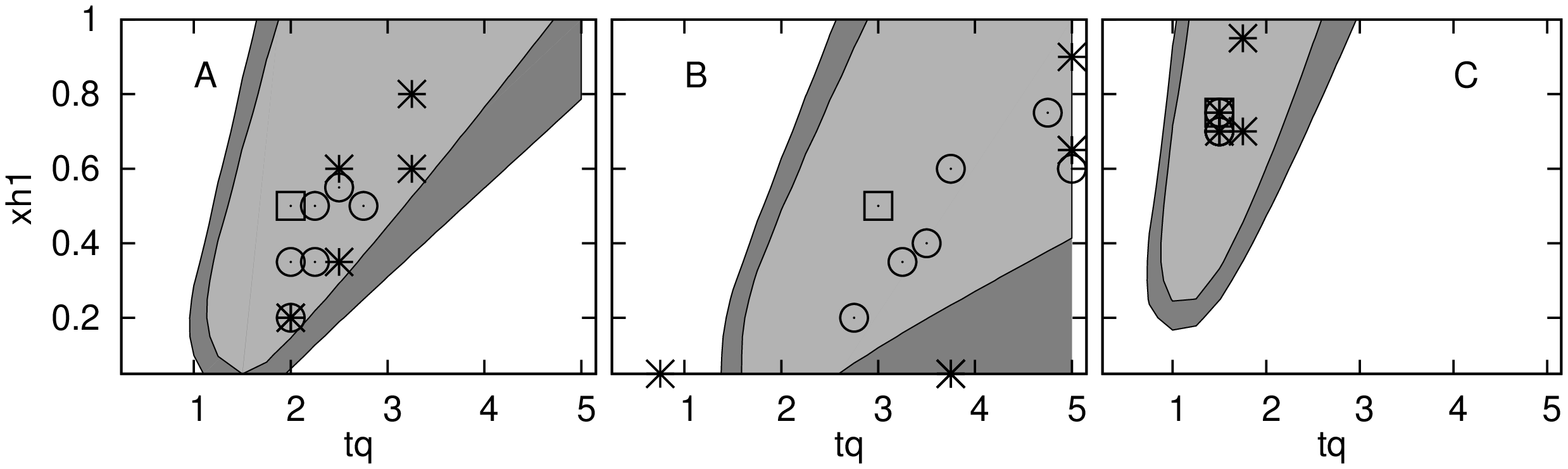}
\includegraphics[width=.3\textwidth, angle=-90]{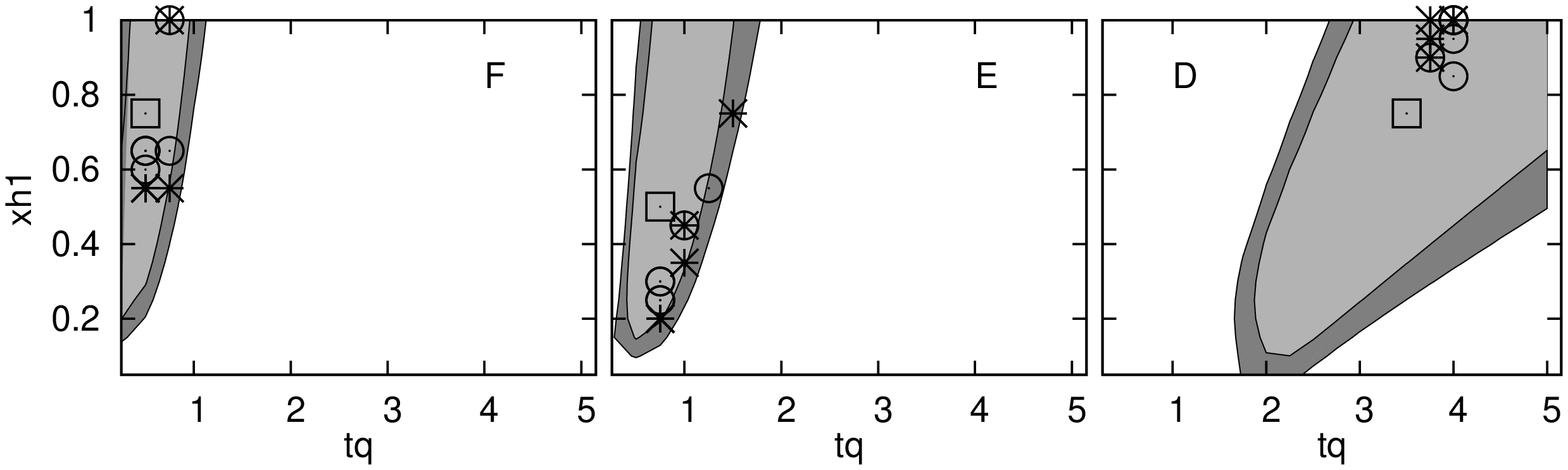}
\caption{This shows the location of the peak SNR for the different
  sets of parameters in Table \ref{tab:sim}.  The square in each panel
  indicates the actual value of $\xh1$ and $\tau_Q$, the circles and
  stars correspond to the estimated values from different realizations
  of the simulations with to uniform and non-uniform recombination
  respectively. The shaded regions correspond to the analytic
  estimates of the uncertainty ($\Delta$SNR $=1$) for $4,000$ and
  $9,000$ hr of observations.}
\label{fig:ani_filter_search}
\end{figure*} 

Distortions in the \HII bubble and fluctuations in the \HI
distribution outside the bubble will affect the SNR for bubble
detection and also the radius $R_{\perp}$ estimated from the best
matched filter. Figure \ref{fig:SNR} shows a comparison of the expected
SNR from our analytic estimates and the SNR obtained from  the
simulations.  We note that the expected SNR has values $2.46$ and
$2.26$ for the parameters sets E and F, and a $3\sigma$ detection is
not possible with $1,000$ hr of observation. We need $1,500$ and
$1,750$ hr of observation for a $3\sigma$ detection for E and F
respectively.

We find that the SNR is below the expected value in cases A and B,
whereas it is comparable to the expected value in D and E, and it is
larger than the expected value in C and F.  We note that the reduction
in SNR seen in A and B may be an artifact of the limited volume of our
simulation where the bubble becomes  comparable to the box size. 
The sets C and F where the SNR is larger than the expected value 
 both correspond to an  early stage of the growth in an highly neutral
IGM ($\xh1 = 0.75$).
 We also find that 
there is a significant scatter in the SNR values amongst the different
realizations. This scatter is relatively smaller for the sets which
have a higher neutral fraction $\xh1 = 0.75$ where we expect the halos
to make a relatively smaller contribution to the ionizing photons. The
scatter in the SNR remains the same even when we consider non-uniform
recombination. However, the overall values of the SNR obtained from
the different realizations are now slightly reduced with respect to
uniform recombination. The fact that the SNR goes down if we introduce
non-uniform recombination is a consequence of the lower contrast
between the ionized bubble and the IGM outside. 

Figure \ref{fig:rperp} shows the bubble radius $R_{\perp}$ estimated
from the best matched filter for the simulations compared with the
analytic estimates. As discussed earlier, we find that the simulated
bubble is nearly always larger than what is expected from
estimates. This discrepancy is particularly noticeable ($\sim 15 - 20
\%$) in the cases which have a lower neutral fraction $\xh1 = 0.5$ and
where we expect the halos to make a relatively larger
contribution. Further as noted earlier the discrepancy is larger in
the simulations with non-uniform recombination in comparison to those
with uniform recombination.
    
We next use the simulated \HII bubbles to analyze parameter
estimation.  Figure \ref{fig:ani_filter_search} shows the parameter
values corresponding to the peak SNR for the different realizations of
the simulations. We see that the values of $\tau_Q$ and $\xh1$
recovered from the simulations are different from the actual age and
neutral fraction. Reason for these deviations are the distortions in
the bubble due to the other ionizing sources and the inhomogeneities
in the IGM.  Figure \ref{fig:ani_filter_search} also shows analytic
predictions of the uncertainties ($\Delta$SNR $=1$) in the estimated
values of $\xh1$ and $\tau_Q$ for the different sets (Table
\ref{tab:sim}) considering $4,000$ and $9,000$ hr of observation. We
have not shown $1,000$ hr observation as bubble detection is not
possible in this observation time for a few of the cases (E and F)
that we have considered. We find that most of the simulations give
parameter estimates which lie within the region corresponding to
$\Delta$SNR $=1$ for $9,000$ hr of observation. We first discuss the
sets E and F where the bubble is at a very early stage of its growth
and is rather small. We require more than $1,000$ hr of observation
for a $3 \sigma$ detection. We find that in these two sets the neutral
fraction is underestimated in almost all the realizations. This is a
consequence of the fact that the actual bubble is almost always larger
than expected (Figure \ref{fig:rperp}) due to the presence other
ionizing sources within the bubble. This effect is more pronounced when
the IGM neutral fraction is low ($\xh1 = 0.5$ in set E). This
discrepancy in the bubble size only affects the estimated neutral
fraction. The bubble in its early stage is very sensitive to $\tau_Q$
through the anisotropy $\eta$ and the shift parameter $s$ (Figure
\ref{fig:contour_eta_s}) which are not severely affected by the
overall size of the bubble. We find that in sets E and F the quasar
age inferred from the simulations is quite close to the actual value.
 
We next consider A, B, C and D which correspond to later stages of
the bubble's growth. We find that the scatter in the values of $\xh1$
and $\tau_Q$ inferred from the simulations is considerably smaller for
a high neutral fraction $\xh1 = 0.75$ (C and D) as compared to $\xh1 =
0.5$ (A and B) where the halos make a relatively larger contribution
to the ionizing photons.  In sets A and B the estimated $\xh1$ is
found to be uniformly scattered around the actual value. However, the
value of $\tau_Q$ is overestimated in the majority of the simulations.
In set D we find that both $\xh1$ and $\tau_Q$ are slightly
overestimated in the simulations, while nearly all the simulated
values lie very close to the actual value in set C.

In summary, our analysis is based on the assumption that the SNR of
the estimator will peak when the parameters of the filter exactly
match those of the bubble that is actually present in the data. The
statistical uncertainty of the estimator will, however, cause the peak
to shift introducing an uncertainty in parameter estimation. Further,
distortions in the shape and size of the bubble due to other ionizing
sources and the inhomogeneities in the IGM will also cause the peak
SNR to shift, introducing further uncertainties in parameter
estimation. In our work we have used the criteria $\Delta$SNR $=1$ to
analytically estimate the statistical uncertainties.  Further, we have
used simulations to assess the effect of the distortions. We find that
the statistical uncertainties are large in most of the situations
that we have considered, and hence these can be used to determine the
uncertainties in the estimated parameter. We find that reliable
parameter estimation is possible using \HII bubbles for which a $3
\sigma$ detection is possible in $1,000$ hr of observation.  Smaller
bubbles, which require longer observations, do not provide very
reliable estimates of $\xh1$, they can however be used to obtain a
reliable estimate of $\tau_Q$.

At the moment we do not have a rigorous justification for the criteria
$\Delta$SNR $=1$ which we have adopted here. A more detailed
statistical analysis involving extensive simulations with large number
of realizations of both the statistical fluctuations as well as the
distortions are required. At the moment this is beyond the
computational power available to us. However, 21-cm observations of
ionized bubbles hold the promise of allowing us to probe the age of
the quasar and the neutral fraction of the IGM, and we plan to pursue
such simulations in future.

\section{acknowledgments}
We would like to thank the anonymous reviewer for providing
us with constructive comments and suggestions which helped to improve
the paper. Suman Majumdar would like to thank Kanan Kumar Datta for
useful discussions and programing related helps during this
work. Suman Majumdar would like to acknowledge Council of Scientific
and Industrial Research (CSIR), India for providing financial
assistance through a senior research fellowship (File No. 9/81
(1099)/10-EMR-I). Suman Majumdar would also like to thank
Harish-Chandra Research Institute, Allahabad, for their warm
hospitality during a visit in February, 2011, when this work was
initiated.


\begin{thebibliography}{99}

\bibitem[\protect\citeauthoryear{Ali, Bharadwaj \&
    Chengalur}{2008}]{ali08} Ali, S.~S., Bharadwaj, S., \& Chengalur,
  J.~N.\ 2008, \mnras, 385, 2166

\bibitem[\protect\citeauthoryear{Bernardi et al.}{2009}]{bernardi09}
  Bernardi, G., de Bruyn, A.~G., Brentjens, M.~A., et al.\ 2009, \aap,
  500, 965

\bibitem[\protect\citeauthoryear{Bharadwaj \&
    Srikant}{2004}]{Bharadwaj04} Bharadwaj, S., \& Srikant,
  P.~S.\ 2004, Journal of Astrophysics and Astronomy, 25, 67

 
\bibitem[\protect\citeauthoryear{Bolton et al.}{2011}]{bolton11}
  Bolton, J.~S., Haehnelt, M.~G., Warren, S.~J., Heweet, P.~C.,
  Mortlock, D.~J., Venemans, B.~P., McMahon, R.~G., Simpson,
  C. \ 2011, \mnras, 416, L70

\bibitem[\protect\citeauthoryear{{Choudhury} \&
  {Ferrara}}{2006}]{choudhury06}
{Choudhury}, T.~R.,  {Ferrara}, A., 2006, Cosmic Polarization, 
Editor - R. Fabbri(Research Signpost), p. 205, arXiv:astro-ph/0603149


\bibitem[\protect\citeauthoryear{Choudhury}{2009}]{choudhury09a}
Choudhury, T.~R.,  2009, Current Science, 97, 6, 841

\bibitem[\protect\citeauthoryear{Choudhury, Haehnelt \&
    Regan}{2009}]{choudhury09b} Choudhury, T.~R., Haehnelt, M.~G., \&
  Regan, J.\ 2009, \mnras, 394, 960


\bibitem[\protect\citeauthoryear{Datta, Bharadwaj \& Choudhury}{2007}]{datta2}
Datta, K.~K., Bharadwaj, S., \& Choudhury, T.~R., 2007,\mnras, 382, 109  

\bibitem[\protect\citeauthoryear{Datta et al.}{2008}]{datta3} Datta,
  K.~K., Majumdar, S., Bharadwaj, S., \& Choudhury, T.~R.,
  2008,\mnras, 391, 1900

\bibitem[\protect\citeauthoryear{Datta, Bharadwaj \&
    Choudhury}{2009}]{datta09} Datta, K.~K., Bharadwaj, S., \&
  Choudhury, T.~R.\ 2009, \mnras, 399, L132

\bibitem[\protect\citeauthoryear{Datta et al.}{2012}]{datta11} Datta,
  K.~K., Mellema, G., Mao, Y., Iliev, I. T., Shapiro, P. R., Ahn,
  K.\ 2012, \mnras, 424, 1877

\bibitem[\protect\citeauthoryear{Datta et al.}{2012}]{datta12} Datta,
  K.~K., Friedrich, M.~M., Mellema, G., Iliev, I.~T., \& Shapiro,
  P.~R.\ 2012, \mnras, 424, 762 

\bibitem[\protect\citeauthoryear{Davis et al.}{1985}]{davis} 
Davis,  M., Efstathiou, G., Frenk, C.~S., \& White, S.~D.~M.
\ 1985, \apj, 292, 371 

\bibitem[\protect\citeauthoryear{Fan et al.}{2003}]{fan03} Fan, X.,
  Strauss, M.~A., Schneider, D.~P., et al.\ 2003, \aj, 125, 1649

\bibitem[\protect\citeauthoryear{Furlanetto, Zaldarriaga \& Hernquist}
  {2004}]{furlanetto1} {Furlanetto}, S.~R., {Zaldarriaga}, M. \&
  {Hernquist}, L. 2004, \apj, 613, 1


\bibitem[\protect\citeauthoryear{Geil et al.}{2008}]{geil2} 
Geil, P.~M., Wyithe, J.~S.~B., Petrovic, N., \& Oh, S.~P.\ 2008,
\mnras, 390, 1496

\bibitem[\protect\citeauthoryear{Gnedin \& Ostriker}{1997}]{gnedin97}
  Gnedin, N.~Y., \& Ostriker, J.~P.\ 1997, \apj, 486, 581


\bibitem[\protect\citeauthoryear{Haehnelt et al.}{1998}]{haehnelt98}
  Haehnelt, M.~G., Natarajan, P., \& Rees, M.~J.\ 1998, \mnras, 300,
  817

\bibitem[\protect\citeauthoryear{Haiman \& Hui}{2001}]{haiman01}
  Haiman, Z., \& Hui, L.\ 2001, \apj, 547, 27


\bibitem[\protect\citeauthoryear{Jarosik et al.}{2011}]{jarosik}
  Jarosik, N. et al.\ 2011, \apjs, 192, 14

\bibitem[\protect\citeauthoryear{Jenkins et al.}{2001}]{jenkins}
  Jenkins, A., Frenk, C.~S., White, S.~D.~M., Colberg, J.~M., Cole,
  S., Evrard, A.~E., Couchman, H.~M.~P., \& Yoshida, N.\ 2001, \mnras,
  321, 372

\bibitem[\protect\citeauthoryear{Komatsu et al.}{2011}]{komatsu}
  Komatsu, E. et al.\ 2011, \apjs, 192, 18


\bibitem[\protect\citeauthoryear{Lu \& Yu}{2011}]{lu11}
  Lu, Y., \& Yu, Q.\ 2011, \apj, 736, 49

\bibitem[\protect\citeauthoryear{Majumdar et al.}{2011}]{majumdar11}
  Majumdar, S., Bharadwaj, S., Datta, K.,~K., \& Choudhury,
  T.~R.\ 2011, \mnras, 413, 1409

\bibitem[\protect\citeauthoryear{Maselli et al.}{2007}]{maselli07}
  Maselli, A., Gallerani,  S., Ferrara, A., \& Choudhury, T.~R.\ 2007,
  \mnras, 376, L34  

\bibitem[\protect\citeauthoryear{Mitra et al.}{2011}]{mitra} Mitra,
  S., Choudhury, T.~R., \& Ferrara, A.\ 2011, \mnras, 413, 1569

\bibitem[\protect\citeauthoryear{Mortlock et al.}{2011}]{mortlock11}
  Mortlock, D.,~J., et al.\  2011, \nat, 474, 7353


\bibitem[\protect\citeauthoryear{Sethi \& Haiman}{2008}]{sethi08} 
Sethi, S., \& Haiman, Z.\ 2008, AJ, 673, 1S


\bibitem[\protect\citeauthoryear{Shapiro \& Giroux}{1987}]{shapiro87} 
Shapiro, P. R., \& Giroux, M. L.\ 1987, \apl, 321L, 107S

\bibitem[\protect\citeauthoryear{Sheth \& Tormen}{2002}]{sheth} 
Sheth, R.~K., \& Tormen, G.\ 2002, \mnras, 329, 61 

\bibitem [\protect\citeauthoryear{Swarup et al.}{1991}]{swarup}  
Swarup G., Ananthakrishnan S., Kapahi V.K., Rao A.P., Subramanya
C.R., Kulkarni V.K.,1991 Curr.Sci.,60,95

\bibitem[\protect\citeauthoryear{White, Becker, Fan \& Strauss}{2003}]
  {white03} White, R. L., Becker, R. H., Fan, X., \& Strauss,
  M. A.\ 2003, AJ, 126, 1

\bibitem[\protect\citeauthoryear{Willott et al.}{2007}]{willott07}
  Willott, C.~J., Delorme, P., Omont, A., et al.\ 2007, \aj, 134, 2435

\bibitem[\protect\citeauthoryear{Willott et al.}{2010}]{willott10}
  Willott, C.~J., Albert, L., Arzoumanian, D., et al.\ 2010, \aj, 140,
  546

\bibitem[\protect\citeauthoryear{Worseck \&
    Wisotzki}{2006}]{worseck06} Worseck, G., \& Wisotzki, L.\ 2006,
  \aap, 450, 495

\bibitem[\protect\citeauthoryear{Wyithe \& Loeb}{2004}]{wyithe04}
Wyithe, J.~S.~B., \& Loeb, A.\ 2004, \apj, 610, 117

\bibitem[\protect\citeauthoryear{Wyithe, Loeb \& Barnes}{2005}]{wyithe05} 
Wyithe, J.~S.~B., Loeb, A., \& Barnes, D.~G.\ 2005, \apj, 634, 715 

\bibitem[\protect\citeauthoryear{Yu \& Lu}{2005}]{yu05b} 
Yu, Q. \& Lu, Y. \ 2005, \apj, 620, 31

\bibitem[\protect\citeauthoryear{Yu}{2005}]{yu05} 
Yu, Q. \ 2005, \apj, 623, 683






\end{thebibliography}
\end{document}